\documentclass[twocolumn]{svjour3}          
\smartqed  
\usepackage{graphicx}
\usepackage{latexsym}
\usepackage{xcolor}

\usepackage[utf8]{inputenc}

\usepackage{natbib}
\bibpunct{(}{)}{;}{a}{,}{,}

\usepackage[breaklinks]{hyperref}
\hypersetup{
    colorlinks = true,
    allcolors = blue,
}

\newcommand{\ariel}{\textit{Ariel}}
\newcommand{\sun}{_\odot}
\newcommand{\earth}{_\oplus}
\newcommand{\kepler}{\textit{Kepler}}
\newcommand{\emcee}{\texttt{emcee}}

\definecolor{darkcyan}{rgb}{0.0, 0.55, 0.55}
\definecolor{dartmouthgreen}{rgb}{0.05, 0.5, 0.06}
\definecolor{magenta}{rgb}{1., 0., 1.}

\begin{document}

\title{Exploiting the transit timing capabilities of \ariel}

\titlerunning{Transit time with \ariel}        

\author{Luca Borsato \and
        Valerio Nascimbeni \and 
        Giampaolo Piotto \and
        Gyula Szab\'{o}
}


\institute{L. Borsato \at
              INAF-Osservatorio Astronomico di Padova \\
              Vicolo dell'Osservatorio 5, - 35122 - Padova - Italy \\
              \email{luca.borsato@inaf.it}           
           \and
           V. Nascimbeni \at
              INAF-Osservatorio Astronomico di Padova \\
              Vicolo dell'Osservatorio 5, - 35122 - Padova - Italy
           \and
           G. Piotto \at
              Universit\`{a} di Padova \\
              Dipartimento di Fisica e Astronomia ``Galileo Galilei'' \\
              Vicolo dell'Osservatorio 3, 35122 Padova - Italy
           \and
           G. Szab\'{o} \at
               ELTE E\"{o}tv\"{o}s Lor\'{a}nd University \\
               Gothard Astrophysical Observatory\\
               Szent Imre h. u. 112, Szombathely, Hungary
}

\date{Received: 2020 June 30 / Accepted: 2021 March 15}

\maketitle

\begin{abstract}
The Transit Timing Variation (TTV) technique is a powerful dynamical tool to measure exoplanetary masses by analysing transit light curves.
We assessed the transit timing performances of the \ariel{} Fine Guidance Sensors (FGS1/2) based on the simulated light curve of a bright, 55 Cnc, and faint, K2-24, planet-hosting star.
We estimated through a Markov-Chain Monte-Carlo analysis
the transit time uncertainty at the nominal cadence of $1$ second and, as a comparison, at a 30 and 60-s cadence.
We found that at the nominal cadence \ariel{} will be able to measure the transit time with a precision of about 12s and 34s, for a star as bright as 55 Cnc and K2-24, respectively.
We then ran dynamical simulations, also including the \ariel{} timing errors, 
and we found an
improvement on the measurement of planetary masses
of about 20-30\%
in a K2-24-like planetary system through TTVs.
We also simulated the conditions that allow us to detect the TTV signal
induced by an hypothetical external perturber within the mass range between Earth and Neptune
using 10 transit light curves by \ariel.
\keywords{\ariel \and transit times \and TTV \and planetary dynamics}
\end{abstract}

\section{Introduction}
\label{intro}

The Transit Time Variation (TTV) technique is a powerful tool to discover 
multi-bodies orbiting around a star,
and can be used to
characterise exoplanetary systems by measuring changes in the orbital period
due to gravitational interaction between planets
\citep{Agol2005MNRAS.359..567A, HolmanMurray2005Sci...307.1288H, Miralda-Escude2002ApJ...564.1019M}.
The amplitude of the TTV effect is greatly boosted if the planets are near a mean-motion resonance
\citep[MMR,][]{Agol2005MNRAS.359..567A, HolmanMurray2005Sci...307.1288H},
allowing us to infer, in particular configurations, even the presence of an exoplanet in the Earth-mass regime.
The \textit{Kepler} mission \citep{Borucki2011ApJ...736...19B},
and its extension \textit{Kepler/K2} \citep{Howell2014PASP..126..398H},
demonstrated the potentiality of this technique,
able to characterise planets also
for stars too faint for a spectroscopy analysis and radial velocity measurement.
Some examples of the application of the TTV technique on multiple planet systems
are:
Kepler-9 \citep{Holman2010Sci...330...51H, Borsato2014A&A...571A..38B, Borsato2019MNRAS.484.3233B},
Kepler-11
\citep{Lissauer2011Natur.470...53L, Lissauer2013ApJ...770..131L}, 
Kepler-88 \citep{Nesvorny2013ApJ...777....3N, Weiss2020AJ....159..242W},
K2-24 \citep{Petigura2016ApJ...818...36P, Petigura2018AJ....156...89P},
and WASP-47 \citep{Vanderburg2017AAS...23040204V, Weiss2017AJ....153..265W}.\\
A TTV can be the result of other different physical sources, e.g.
star spots,
oblateness of hosting star,
orbital precession induced by tidal interactions, 
and general relativity \citep[for a general review see][]{Perryman2018exha.book.....P},
and non-physical sources such as binning or sampling of the data \citep[e.g.][]{Kipping2010MNRAS.408.1758K, Mazeh2013ApJS..208...16M, Szabo2013A&A...553A..17S, PriceRogers2014ApJ...794...92P}.
In this work we focus on TTV only due to gravitational interaction of exoplanets.
\par

The European Space Agency (ESA) adopted
\ariel{} \citep[Atmospheric Remote sensing Infrared Exoplanet Large survey;][]{Pascale2018SPIE10698E..0HP, Pilbratt2019ESS.....450304P, Puig2018ExA....46..211P, Tinetti2018ExA....46..135T}
as an M4 mission within the Cosmic Vision programme,
and its launch is scheduled for 2028.
The \ariel{} mission aims to study the atmospheres and the chemical compositions of exoplanets
by surveying about 1000 transiting exoplanets through fast-cadence high-precision spectroscopy
and photometry of transits, eclipses, and phase-curves
in a wide spectral range from
the visible (VIS, 0.5~$\mu$m) to the infrared 
\citep[IR, 7.8~$\mu$m; detailed description in][]{Tinetti2018ExA....46..135T, Encrenaz2018ExA....46...31E, Zingales2018ExA....46...67Z, Pascale2018SPIE10698E..0HP}. 
The \ariel{} satellite mount
two spectrometers,
the \ariel{} Infrared Spectrometer (AIRS, two prism-dispersed channels that cover the band width 1.95--3.9$\mu$m with $R > 100$ and 3.9--7.8$\mu$m with $R > 30$) and
NIRSpec (slit-less prism spectrometer with spectral resolving power $R > 15$ in spectral range of 1.1--1.95~$\mu$m),
and three photometers,
namely VISPhot (1.1--1.95~$\mu$m) and
two \ariel{} Fine Guidance Sensors (FGS1 at 0.6--0.8~$\mu$m and FGS2 at 0.8--1.11~$\mu$m).
See \citet{Tinetti2018ExA....46..135T}, \citet{Pascale2018SPIE10698E..0HP},
and \citet{Mugnai2020ExA....50..303M} for further details on \ariel{} instruments.
The fast-cadence photometric data
will allow us to measure the planetary transit time
at precision and accuracy level of a few seconds.
This is crucial to extend the temporal baseline of TTV signals
of known planets in multiple-planet systems,
breaking the degeneracy on doubtful dynamical solutions
and improving the precision on their masses and orbital parameters
\citep{Petigura2018AJ....156...89P,Delrez2018MNRAS.475.3577D}.
\par

In this work we present a study on the \ariel{} timing performances
based on simulations of real targets (Sec.~\ref{sec:2}) and some possible science cases (Sec.~\ref{sec:3}).
This analysis describes one of the possible extended uses of the \ariel's Core data and
it has been developed within the \ariel{} Working Group High-Precision Photometry
(described in details in Szabo et al.; 2020, submitted and
in Haswell 2020, in preparation).
\par

\section{Transit Time analysis}
\label{sec:2}

We simulated two science cases, modelled after the real planet hosts 55 Cnc e and K2-24 b,
by analysing synthetic light curves assuming the nominal performances of FGS1 and 2.
In our analysis we simultaneously fitted the light curves gathered from the two channels
to get a single value of the parameters that are not dependent on wavelength, such as the $T_0$.
We simulated photometric data at three different cadences: 1~s,
that is the nominal cadence of \ariel{} observations,
30 and 60~s as a comparison, in order
to understand when the sampling cadence becomes a limiting factor for our analysis.\par

For both targets we estimated the expected \ariel{} photometric noise per time unit,
and for each band-pass, by using the ArielRad code 
\citep{Mugnai2020ExA....50..303M}.
This program takes into account different sources of noise,
such as detector noise (dark current, gain and read-out noise),
and other stationary random processes modelled as Poisson noise,
i.e. photon noise, Zodiacal noise, and instrument emission).
ArielRad cannot include non-stationary noise (time-correlated noise), so
it includes a jitter noise as the variance at different wavelength on timescale of 1 hour,
and it includes a margin noise to take into account noise uncertainties and possible time-correlated effects
\citep[about 20\% on top of the photon noise, see][]{Mugnai2020ExA....50..303M}.
We re-scaled the noise to the actual sampling cadence and
we added a noise floor of 20~ppm in quadrature 
\citep[as shown in eq.~17 of ][]{Mugnai2020ExA....50..303M}.
This noise floor is needed to avoid to indefinitely integrate down the noise \citep{Beichman2014PASP..126.1134B, Greene2017AAS...23030901G, Molliere2017A&A...600A..10M}, and
it is 20~ppm at every integration time 
\citep{Mugnai2020ExA....50..303M}.
Due to the lack of time-correlated model in the noise analysis of ArielRad,
we also added a linear trend (as $0.2\%$ on the transit time scale) to the light curve
to mimic the presence of systematic effects from astrophysical (e.g. stellar activity)
or from instrumental sources (e.g. due to stray light or pointing effects of the satellite).
We scaled the FGS2 trend as a random Gaussian with $\sigma = 0.005)$.
We also created light curves with a quadratic term\footnote{We used coefficients $c_2=-0.003$, $c_1=0.002$, $c_0=0.0$,
from high to low order.},
and scaled the FGS2 as before.
We analysed the statistics of Table~4 of \citet{Holczer2016ApJS..225....9H} and
we found that for 92.2\% of the planets (on a sample of 2339)
the trend could be described by a polynomial of first order;
the rest can be detrendend with a second and third order polynomial (4.7\% and 3.2\%, respectively).
So, we can assume that in most cases a linear trend is a good approximation of the out-of-transit on
timescales of 2--3 transit duration.
However, the measurement of the transit time and of its uncertainty
is strongly affected by the sampling of the ingress and egress phases
and by the symmetry of the transit model,
so, the trend (and the red noise) contributes only at a second order.
We decided to only fit a linear trend even if the light curve has been created with a quadratic one
and a red noise term.
\par

We modelled the transit light curve with the \texttt{batman} package \citep{Kreidberg2015PASP..127.1161K}
and we ran a statistical analysis with the 
\emcee{}\footnote{
We are aware that \emcee{} code should be used with caution when space of the parameters has a number of 
dimension greater than 10,
but it has been extensively used in exoplanet literature showing great performances in different cases.
Furthermore, using a parameterization that reduces the correlation among parameters and
that allows the algorithm to evenly sample the parameter space
can mitigate this issue.
}
package
\citep{DFM2013ascl.soft03002F, DFM2019JOSS....4.1864F},
a Markov-Chain Monte-Carlo (MCMC) code with an affine-invariant sampler 
\citep{GoodmanWeare2010CAMCS...5...65G}.
\par

Firstly, we generated the synthetic transit light curve with the physical parameters reported in Table~\ref{tab:1}.
Then, for each cadence, we fitted
common parameters 
(different parameterization with respect to Table~\ref{tab:1})
between the two filters, i.e.
the stellar density in solar units ($\rho_\star$),
the base-2 logarithm of the period ($\log_2 P$, used in combination with $\rho_\star$ to constraint the $a/R_\star$),
the impact parameter\footnote{
In the impact parameter we taken into account the eccentricity and the argument of pericenter as in \citet{Winn2010exop.book...55W} and \citet{Kipping2010MNRAS.407..301K}.
}
($b$),
a combination of eccentricity $e$ and argument of pericenter $\omega$
($\sqrt{e}\cos\omega$, $\sqrt{e}\sin\omega$),
and the transit time ($T_0$),
while, for each filter, we fitted 
the planet to stellar radius ratio ($k = R_\mathrm{p}/R_\star$),
the quadratic limb-darkening (LD) coefficients, $q_1$ and $q_2$, 
an uninformative sampling of quadratic LD $u_1$ and $u_2$ in the form 
$q_1 = (u_1+u_2)^{2}$ and
$q_2 = 0.5 u_1 (u_1 + u_2)^{-1}$
as suggested in \citet{Kipping2013MNRAS.435.2152K},
a jitter factor in base-2 logarithm ($\log_2 \sigma_\mathrm{jit}$),
and linear trend.
We selected the best (and common) parameterization of the fitting model
to reduce the correlation among parameters and 
allowing the \emcee{} code to properly, and as evenly as possible, sample the parameter space.
We wanted to simulate an \ariel{} transit observation, i.e. different transit depth for each different band-pass
depending on different planetary radius that corresponds to different layers of atmosphere.
Given that the two FGS channels have different, almost non-overlapping,
band-pass \citep[see Figure 2 in ][]{Mugnai2020ExA....50..303M},
we decided to fit different $k$,
but we generated the synthetic light curves with a single value.
We used as priors the stellar density (computed from stellar mass and radius), period $P$
, and
eccentricity reported in Table~\ref{tab:1}, and we set very tight prior if the parameter has a null error (i.e. eccentricity of 55 Cnc e).
We did not apply any priors on the LD coefficients.
\par

The \emcee{} code outputs the posterior distribution of the fitted parameters
from which we extract a best-fit solution 
and an uncertainty as
the high density interval (HDI at $68.27\%$, equivalent of 1-$\sigma$)
for the transit model parameters, including the most relevant here,
i.e. the transit time ($T_0$) and its uncertainty ($\sigma_{T_0}$).
As best-fit solution we compared the median,
the mode\footnote{As $\mathrm{mode} = 3\times \mathrm{median} - 2\times \mathrm{mean}$},
and the maximum likelihood estimation
(MLE, as the parameter sample that maximises the likelihood within the HDI).
In our cases the differences among these best-fit parameter sets are negligible
(in terms of values and uncertainties) and
we decided to use as representative of the best-fit solution
the median of the posterior distribution,
because of its symmetry with respect to the HDI.
We want also to stress that finding the best-fit solution was out of the purpose of this work
and we wanted to focus on the determination of the uncertainty of the parameters,
mainly the $\sigma_{T_0}$.
\par

\subsection{55 Cnc e}
\label{sec:2.1}

The first case, 55~Cnc e, has been selected to assess the performances of \ariel{}
on transit timing at the bright end.
The magnitude and spectral type of the host star 55~Cnc
\citep[V = 5.95~mag, $M_\star = 0.91\, M\sun$, $R_\star = 0.94\, R\sun$;][]{vonBraun2011ApJ...740...49V}
and the size of planet e, $R_\mathrm{e} \sim 2\, R\earth$ \citep{Demory2016Natur.532..207D}
represent a perfect case to assess the feasibility of a solar-like star with a small transiting planet.
Its transit has very short, and almost vertical, ingress and egress phases,
reducing the impact of the LD effect to the timing of the transit model.
\par
We obtained stellar and planetary parameters from 
\citet{vonBraun2011ApJ...740...49V} and \citet{Demory2016Natur.532..207D} and
available through the NASA Exoplanet Archive\footnote{https://exoplanetarchive.ipac.caltech.edu/}.
We computed the coefficients of a quadratic LD-law ($u_1$ and $u_2$) fitting 
the ATLAS \citep{Kurucz1979ApJS...40....1K} and PHOENIX \citep{Husser2013A&A...553A...6H} models 
through the tool\footnote{http://www.github.com/nespinoza/limb-darkening} developed by \citet{Espinoza2015MNRAS.450.1879E}.
We assumed a box-shaped filter for both FGS1 and FGS2 
within the spectral range of $0.8-1.0\,\mu$m and $1.05-1.2\,\mu$m, respectively.
We adopted the values of $u_1$ and $u_2$ for the fitting as the average of the values computed for each combination of stellar parameters and their extremal errors for both models, obtaining: 
$u_1 = 0.33 \,, u_2 = 0.23$ for the FGS1 and
$u_1 = 0.26\,, u_2 = 0.25$ for the FGS2.
We also tested LD coefficients for band $I$ and $J$ computed with \texttt{exofast}\footnote{http://astroutils.astronomy.ohio-state.edu/exofast/limbdark.shtml} \citep{Eastman2013PASP..125...83E}:
$u_1 = 0.39 \,, u_2 = 0.22$ for the $I$ band and
$u_1 = 0.22\,, u_2 = 0.29$ for the $J$ band.
We used the LD coefficients of $I$ for FGS1 and those of $J$ for FGS2 and
we found no difference with the results obtained from the LD coefficients 
computed from the stellar models.
See in Table~\ref{tab:1} the summary of the stellar and planetary parameters used to create the synthetic light curves.\par

\begin{table*}
    \caption{Stellar and planetary parameters adopted in the synthetic light curves. 55 Cnc e parameters from \citet{vonBraun2011ApJ...740...49V,Demory2016Natur.532..207D}, K2-24 b from \citet{Petigura2016ApJ...818...36P, Petigura2018AJ....156...89P}.}
    \label{tab:1}
    \begin{center}
    \begin{tabular}{lll}
        \hline\noalign{\smallskip}
        Parameters & 55 Cnc e & K2-24 b  \\
        \noalign{\smallskip}\hline\noalign{\smallskip}
        $T_\mathrm{eff}\, (\mathrm{K})$ & $5196 \pm 24$ & $5625 \pm 60 $ \\
        $\log g\, (\mathrm{dex})$ & $4.45 \pm 0.01$ & $4.29 \pm 0.05$ \\
        $[Fe/H]\, (\mathrm{dex})$ & $0.31 \pm 0.04$ & $0.34 \pm 0.04$ \\
        $M_\star\, (M\sun)$ & $0.91 \pm 0.02$ & $1.07 \pm 0.06$ \\
        $R_\star\, (R\sun)$ & $0.94 \pm 0.01$ & $1.16 \pm 0.04$ \\
        $R_\mathrm{p}\, (R\earth)$ & $1.91$ & $5.4$ \\
        $P\, (\mathrm{days})$ & $0.736539 \pm 0.000007$ & $20.88977 \pm 0.00035$ \\
        $T_0^{(a)}\, (\mathrm{days})$ & $5733.013$ & $6905.886$ \\
        $e$ & $0$ & $0.06 \pm 0.01$ \\
        $\omega\, (^\circ)$ & $90$ & $90$ \\
        $i\, (^\circ)$ & $83.3$ & $89.25$ \\
        FGS1 &  & \\
        $u_1$ & $0.33$ & $0.29$ \\
        $u_2$ & $0.23$ & $0.25$ \\
        noise (ppm/hr)$^{(b)}$ & $12.4$ & $54.0$ \\
        FGS2 &  & \\
        $u_1$ & $0.26$ & $0.23$ \\
        $u_2$ & $0.25$ & $0.26$ \\
        noise (ppm/hr)$^{(b)}$ & $5.6$ & $43.5$ \\
        \noalign{\smallskip}\hline
    \end{tabular}
    \end{center}
    $^{(a)}$ $T_0$ expressed as $\mathrm{BJD_{TDB}}- 2450000$. 
    $^{(b)}$ Noise value from ArielRad. We added in quadrature a noise floor of 20~ppm as described in the text.
\end{table*}

\begin{figure*}
\centering
\includegraphics[width=0.95\textwidth]{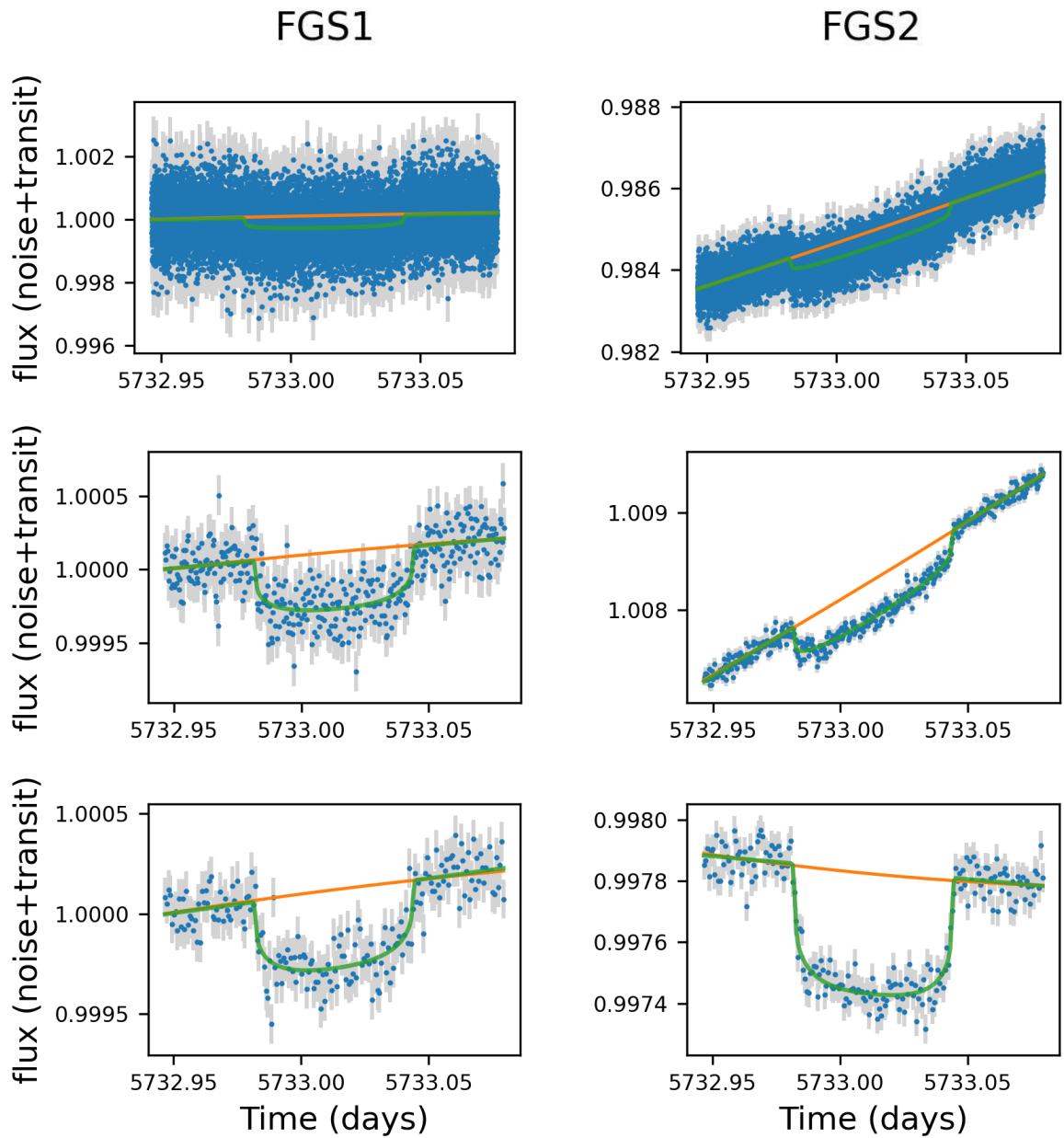}
\caption{Transit light curves of 55 Cnc e with simulated \ariel{} 
noise (blue dots with gray error bars)
and quadratic trend (orange line)
along with the best-fit transit model with fitted linear trend (green line)
for FGS1 (left) and FGS2 (right).
First row for $1$~s cadence, second and third row for $30$~s and $60$~s, respectively.}
\label{fig:1}       
\end{figure*}

The analysis (see best-fit model for each cadence in Fig.~\ref{fig:1}) yields an estimated uncertainty in the transit time ($\sigma_{T_0}$) of
$12$~s, $12$~s, and $14$~s for cadence at $1$~s, $30$~s, and $60$~s respectively,
that is the sampling cadence for bright targets is not a limiting factor.

\subsection{K2-24 b}
\label{sec:2.2}

The second case, K2-24 b, is a Neptune-size transiting planet in a multiple-planet system
showing anti-correlated TTVs with another transiting planet (K2-24 c).
The star is much fainter than 55~Cnc ($V=11.28$, $K = 9.18$) with a mass and radius slightly bigger than the Sun ($M_\star = 1.07\, M\sun$, $R_\star = 1.16\, R\sun$).\par

We studied the transit timing performances of \ariel{} with the same two FGS channels as before and
we also analysed the impact of a few \ariel{} observations on the recovering of the planetary ephemeris and on the improvement of the other orbital parameters (see Sec.~\ref{sec:3}).
We used the stellar parameters from \citet{Petigura2018AJ....156...89P} and 
we computed the quadratic LD coefficients, as we did for 55 Cnc e.
We found
$u_1 = 0.29 \,, u_2 = 0.25$ for the FGS1 and
$u_1 = 0.23\,, u_2 = 0.26$ for the FGS2.
We also tested LD coefficients from \texttt{exofast}:
$u_1 = 0.32 \,, u_2 = 0.27$ for the $I$ band and
$u_1 = 0.17\,, u_2 = 0.30$ for the $J$ band.
A summary of the stellar and planetary parameters are reported in Table~\ref{tab:1}.\par

In the K2-24 b case, when using LD for $I$ and $J$ we obtained
for each cadence the same $\sigma_{T_0}$ of $35$~s.
This result suggests us that in this situation we are photon noise dominated.
In the analysis with the LD from box-shaped FGS1 and 2 filters we obtained
$\sigma_{T_0}$ of $34$~s, $29$~s, and $32$~s for cadence at $1$~s, $30$~s, and $60$~s respectively.
In both analysis the $\sigma_{T_0}$,
obtained with only one transit,
is about half the best timing uncertainty ($\sim 78$~s)
obtained with Kepler/K2 observations
\citep{Petigura2018AJ....156...89P}.
\par

\begin{figure*}
\centering
\includegraphics[width=0.95\textwidth]{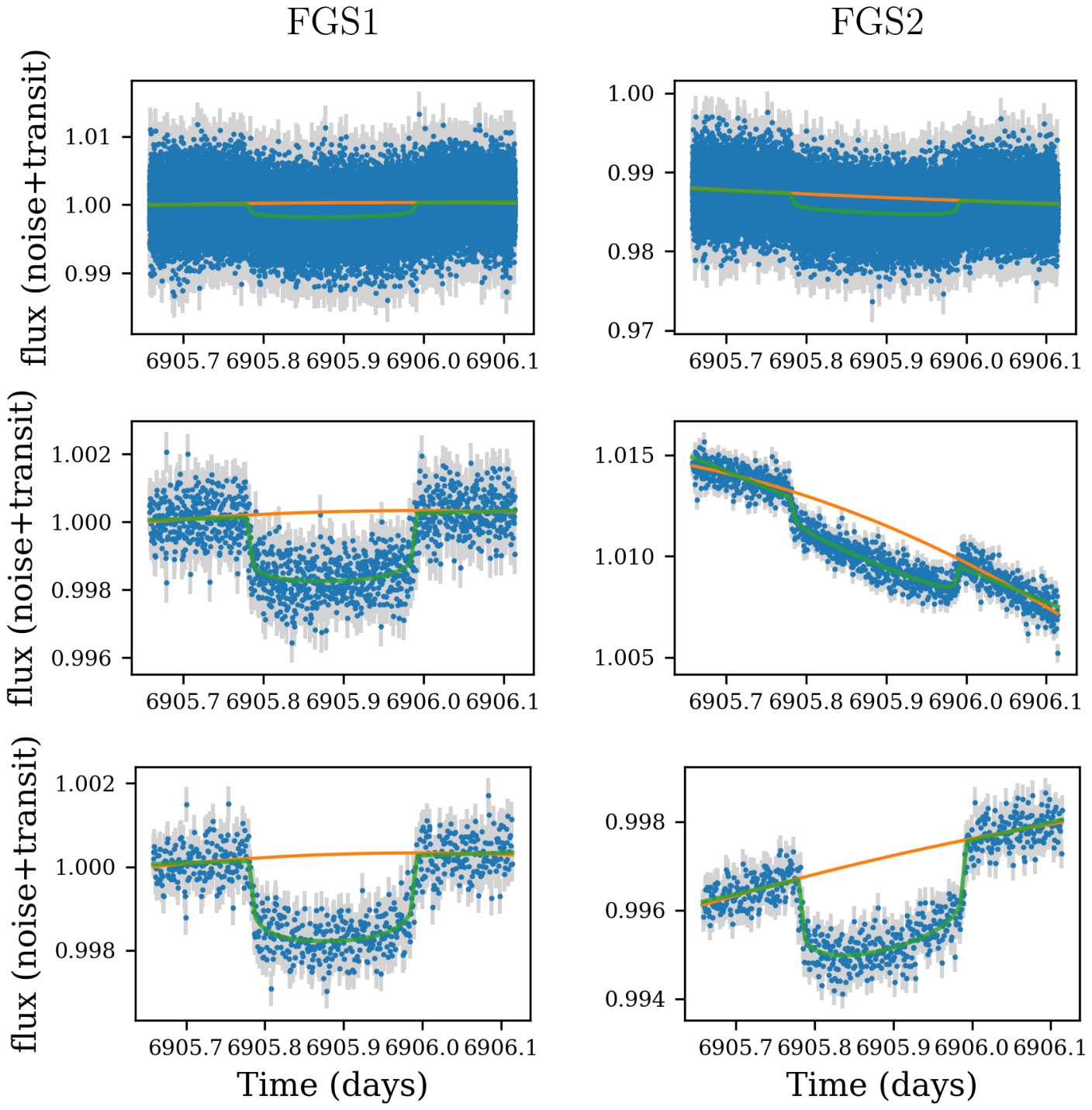}
\caption{Same as in Fig.~\ref{fig:1}, but for K2-24 b.}
\label{fig:2}       
\end{figure*}

We also analysed the case of K2-24 b observed by \ariel{} with only one channel, i.e. the FGS1.
This could mimic the possible failure of one of the instruments,
reducing the number of light curves per transit.
We generated the synthetic light curve with the parameters in Table~\ref{tab:1}
with LD coefficients of FGS1 in $I$ band.
We repeated the analysis as before and
we obtained a $\sigma_{T_0}$ of about 53 seconds,
that is still better than Kepler/K2 timing obtained from the simultaneous analysis
of four transits for this target.
This could be due to the high cadence photometry that is able to homogeneously sample the
ingress and egress, 
the transit phases that have the greatest impact on determining the transit time.
\par

\section{Dynamical analysis}
\label{sec:3}

\subsection{Impact of \ariel{} observations}
\label{sec:3.1}

K2-24 is a multiple-planet system showing anti-correlated TTVs by the transiting planet b and c,
but with a poor sampling of the TTV period on K2 data
\citep{Petigura2016ApJ...818...36P, Petigura2018AJ....156...89P}.
This can lead to an imprecise or 
ambiguous determination of
masses and orbital parameters of the planets
\citep[see the Kepler-9 case by][]{Holman2010Sci...330...51H, Borsato2014A&A...571A..38B, Borsato2019MNRAS.484.3233B}.
\citet{Petigura2018AJ....156...89P} found a possible candidate planet d,
from the radial velocity (RV) analysis, 
with a mass of $54\pm14\, M\earth$ and
a period of about 427 days.
\par

\begin{figure*}
\centering
\includegraphics[width=0.95\textwidth]{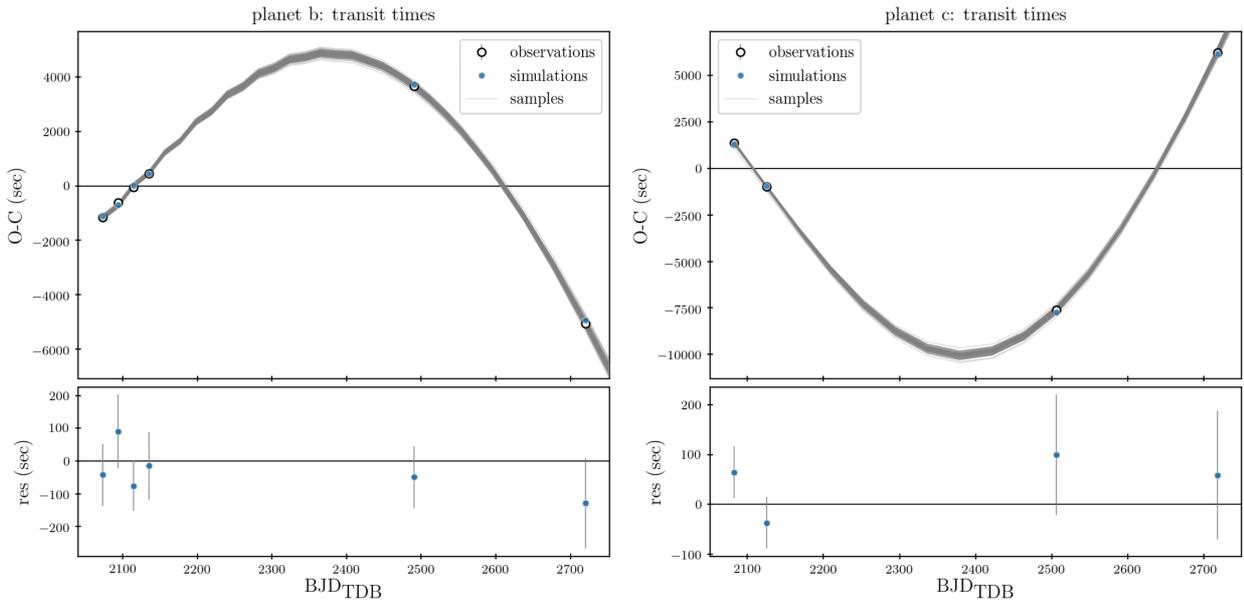}
\caption{Top panel: Observed - Calculated ($O-C$) plot of K2-24 b (left) and c (right)
synthetic transit times for the baseline of the observations.
The Observed is given by the synthetic $T_0$s (open-black dots) or simulated $T_0$s (filled-blue dots),
while Calculated is given by the $T_0$s computed from linear ephemeris (from synthetic $T_0$s).
The gray lines are 100 drawn from the posterior distribution.
Bottom panel: residuals between observed and simulated $O-C$.
The x-axis is in unit of days, starting from a reference time 
$\mathrm{BJD_{TDB}}-2454833 = 2071.0$
to match the transit times from \citet{Petigura2018AJ....156...89P}.
}
\label{fig:3}       
\end{figure*}

We simulates the K2-24 system with the dynamical integrator within the
TRADES code\footnote{https://github.com/lucaborsato/trades}
\citep{Borsato2014A&A...571A..38B, Malavolta2017AJ....153..224M, Borsato2019MNRAS.484.3233B}.
We integrated the orbits for the same time range spanning the observations by Kepler/K2 and follow-up
(radial velocities, RVs, and additional transits) and
computed all the transit times ($T_0$s) and RVs within the same range.
We then selected the corresponding $T_0$s and RVs analysed by \citet{Petigura2018AJ....156...89P} and
we added white noise, based on the actual error bars from the same work.
We used \texttt{TRADES} with the \texttt{emcee} module to get masses and orbital parameters from
this synthetic data set and to compute the so-called ``Observed - Calculated'' ($O-C$) plot, where the residuals of the observed $T_0$ with respect to the reference linear ephemeris are shown as a function of time (see Fig.~\ref{fig:3}).
\par

\begin{figure*}
\centering
\includegraphics[width=0.95\textwidth]{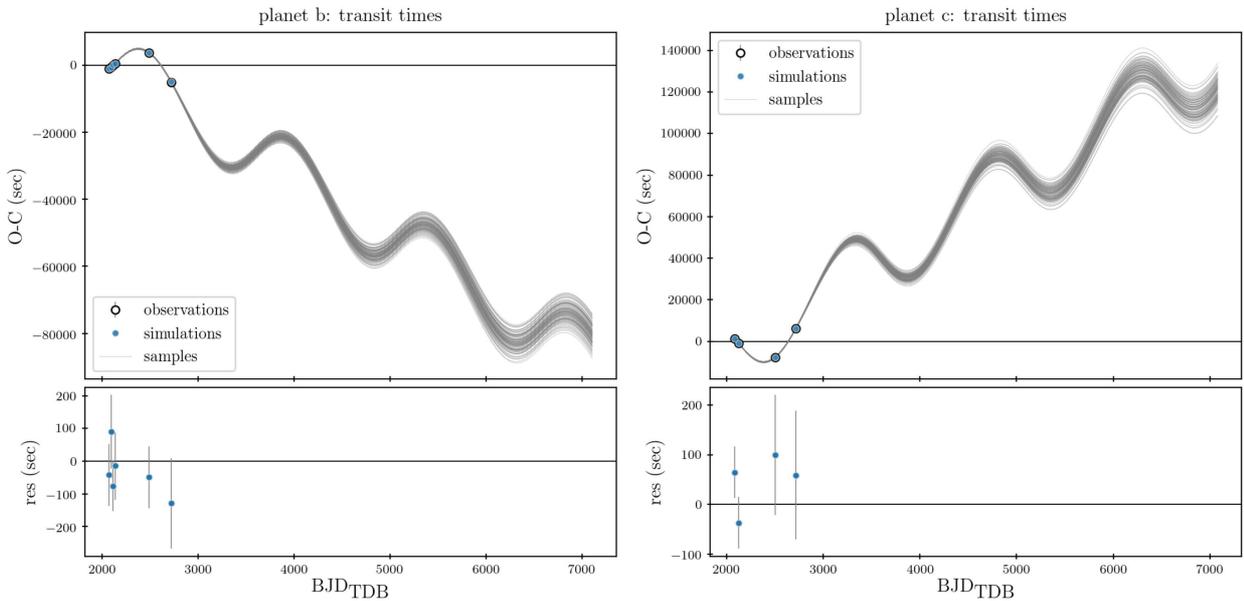}
\caption{Same as Fig.~\ref{fig:3}, but extending the orbital integration to almost 10 years in the future.
The linear trend of the departure from $O-C = 0$ is due to the short time coverage of
the linear ephemeris computed from the synthetic $T_0$s.}
\label{fig:4}       
\end{figure*}

If we integrate for 10 years further the orbital solution obtained,
we have two main effects on the prediction of the transit times.
The accumulating errors of the linear ephemeris produce a shift of the transit time with respect to the predicted one.
Another source of error on the transit time is due to the uncertainty on
the orbital parameters of the planetary system (see. Fig.~\ref{fig:4}).
In 2028, when the \ariel{} launch is supposed to take place,
the shift from the linear ephemeris could lead to almost one day of uncertainty on the transit times.
Even if we could recover the ephemeris,
we still have more than three hours of uncertainty on transit times
due to the errors on the orbital parameters.

\begin{figure*}
\centering
\includegraphics[width=0.95\textwidth]{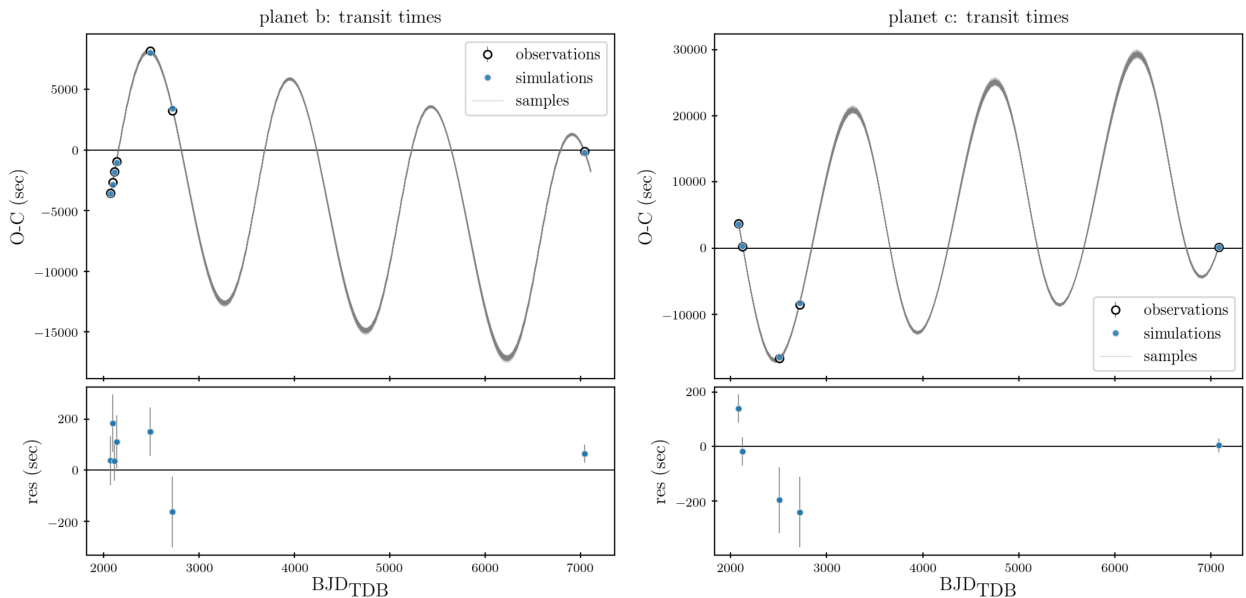}
\caption{Same as Fig.~\ref{fig:4}, but for the analysis with two $T_0$s by \ariel{},
one for planet b (left) and one for planet c (right).
The thin gray area is the result of 100 random draws from the posterior distribution;
this area is thinner than in Fig.~\ref{fig:4} as the result of better determination of orbital parameters.}
\label{fig:5}       
\end{figure*}

We simulated the K2-24 system and ran a dynamical retrieval with \texttt{TRADES+emcee}
in the case of our synthetic data set with two observations with \ariel{},
one for planet b and one for c.
We assumed the same error bar for the two transit times $\sigma_{T_0} = 35$~s,
that is the highest (conservative) value obtained from the previous analysis (see Sec.~\ref{sec:2.2}).
We built the $O-C$ plots (see Fig.~\ref{fig:5}) from the best-fit solution extracted from
the posterior distribution and we found that just one transit per planet is enough
to recover the ephemeris and to greatly improve the knowledge of orbital parameters of the system.
This can be easily achieved taking advantage of existing ground-based transit survey,
such as the
The Asiago Search for Transit timing variations
of Exoplanets project \citep[TASTE,][]{Nascimbeni2011A&A...527A..85N, Granata2014AN....335..797G},
the Next-Generation Transit Survey \citep[NGTS,][]{Wheatley2018MNRAS.475.4476W},
and the ExoClock Project\footnote{https://www.exoclock.space/} (within the \ariel{} Ephemerides Working Group),
or space missions, i.e.
Transiting Exoplanet Survey Satellite \citep[TESS,][]{Ricker2014SPIE.9143E..20R},
CHaracterizing ExOPlanets Satellite \citep[CHEOPS,][]{Broeg2014CoSka..43..498B,Benz2020ExA...tmp...53B},
PLAnetary Transits and Oscillations of stars \citep[PLATO,][]{Rauer2014ExA....38..249R}.
Thanks to TESS and to the extended operations TESS will also be of great advantage 
for the ephemeris recovery of almost all the \kepler/K2 targets on the ecliptic
and most of the targets on the sky.
With only two transit times from \ariel{} we get an improvement
on the uncertainty of the mass of about $22\%$ and $31\%$ for planet b and c, respectively. \par

\subsection{Possible TTV signal from \ariel{} observations}
\label{sec:3.2}

\begin{figure}
\centering
\includegraphics[width=0.95\columnwidth]{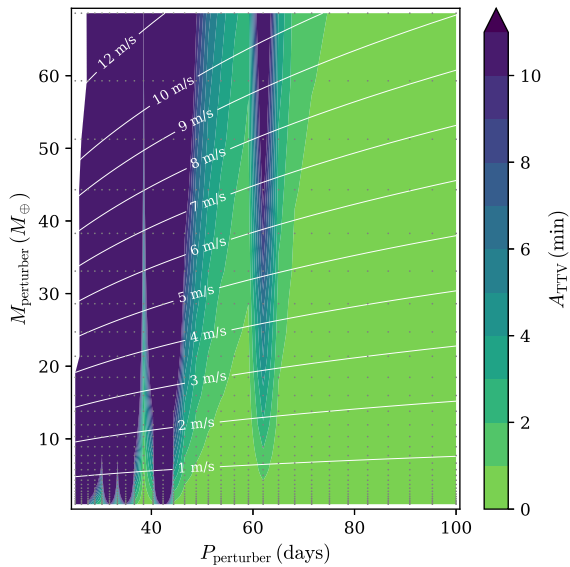}
\caption{TTV amplitude ($A_\mathrm{TTV}$) for K2-24 b as a function of mass and period of the perturber. 
Each gray dot is the combination of mass-period in a grid of 900 simulations,
where $M_\mathrm{perturber} = [1\, M\earth, 4\, M_\mathrm{Nep}]$ and
$P_\mathrm{perturber} = [25, 100]$~days,
both in 30 logarithmic steps.
For each simulation we selected 10 transit times within 4 years of orbital integration,
we repeated the selection 100 times and 
computed the $A_\mathrm{TTV}$ as the the median value of
the semi-amplitude of the $O-C$ with respect to a linear ephemeris.
In these simulations the initial $e_\mathrm{perturber}$ was set to 0.0.
The white contour lines are the RV semi-amplitude due to the perturber.
}
\label{fig:6}       
\end{figure}

\begin{figure}
\centering
\includegraphics[width=0.95\columnwidth]{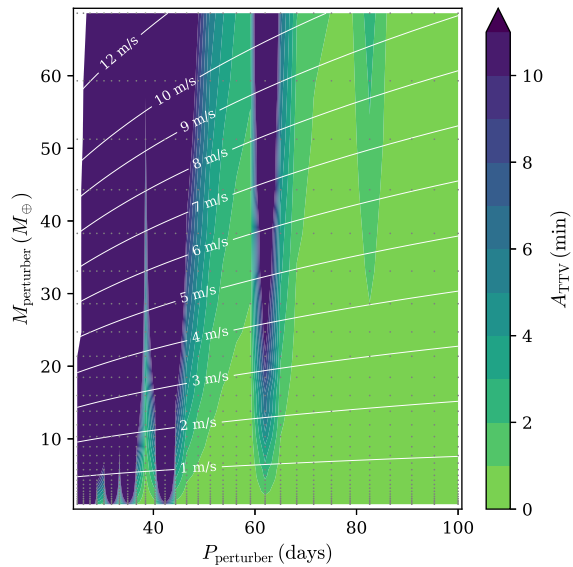}
\caption{Same as Fig.~\ref{fig:6}, but with initial $e_\mathrm{perturber} = 0.05$.
It is evident that a perturber with non-zero eccentricity extend the possible TTV signal regions and increase their amplitudes.}
\label{fig:7}       
\end{figure}

\begin{figure}
\centering
\includegraphics[width=0.95\columnwidth]{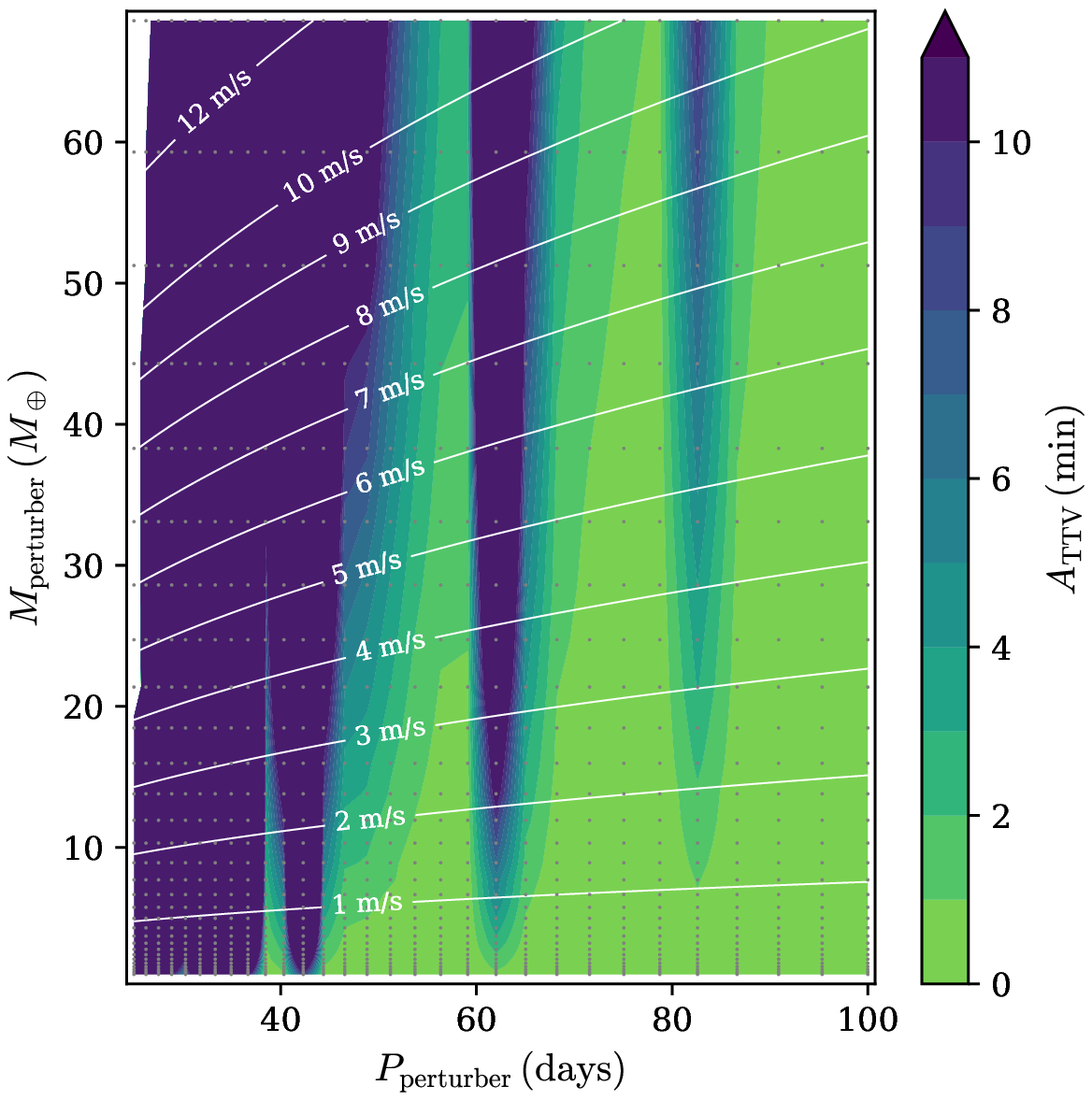}
\caption{Same as Fig.~\ref{fig:6}, but with initial $e_\mathrm{perturber} = 0.1$.
In this case, the effect of the eccentricity of the perturber is even stronger than that in Fig.~\ref{fig:7}.}
\label{fig:8}       
\end{figure}

In the most favourable case \ariel{} will be able to observe 10 transits of the same planet \citep{Puig2018ExA....46..211P, Tinetti2018ExA....46..135T}.
We wanted to determine the amplitude of the TTV signal  ($A_\mathrm{TTV}$) produced by an external
perturber.
We decided to consider as transiting planet the K2-24 b \citep[orbital parameters from ][]{Petigura2016ApJ...818...36P, Petigura2018AJ....156...89P}
and we used \texttt{TRADES},
by choosing the mass and period of an hypothetical outer perturber from a grid
 with 30 log-spaced values of mass,
ranging from $1\, M\earth$ to $4\, M_\mathrm{Nep}$,
and 30 log-spaced values of orbital period,
from 25 to 100 days.
For each simulation the code integrates the orbits for 4 years (as the nominal duration of the \ariel{} mission),
selects the transit times, with associated error of $35$~s determined in Sec.\ref{sec:2.2},
and it computes the linear ephemeris.
Then it selects 10 random transits (without replacement),
re-computes the linear ephemeris,
and calculates the $A_\mathrm{TTV}$ as the semi-amplitude of the $O-C$
(selected transit times minus the newly computed linear ephemeris).
It repeats this for 100 times and computes the median of $A_\mathrm{TTV}$.
We obtained a map of the $A_\mathrm{TTV}$ as a function of the mass ($M_\mathrm{perturber}$)
and of the period ($P_\mathrm{perturber}$).
It is well known that the eccentricity of the perturber ($e_\mathrm{perturber}$) boosts the $A_\mathrm{TTV}$,
so we did the same analysis three times,
with three different initial values of $e_\mathrm{perturber}$: 0.0, 0.05, and 0.1 
(see Figures~\ref{fig:6}, \ref{fig:7}, and \ref{fig:8}).

The darkest regions of the P-M parameter space (shown in Fig.~\ref{fig:5}, \ref{fig:6}, and \ref{fig:7})
where the amplitude $A_\mathrm{TTV}$ is much larger than the timing error $\sigma_{T_0}$
for the individual transit observations ($\sim 35$~s, Sec.~\ref{sec:2.2}) are those where
the TTV signal could be detected at high confidence with \ariel{}.
At least in some orbital configurations close to a MMR, and especially for eccentric orbits,
it will be possible for \ariel{} to robustly detect a TTV signal induced by
an outer pertuber with one Earth mass (or greater)
with just 10 transit observations.

\section{Conclusions}
\label{sec:4}

According to our simulations, the \ariel{} photometric and timing precision is suitable
to study known multiple-planet systems showing TTV signals in order to extend the TTV phase coverage,
breaking the degeneracy on the retrieved orbital parameters, and in particular to improve the mass estimation.
The latter is fundamental to put strong constraints on
planetary formation, migration, and evolution processes \citep[e.g. ][]{Mordasini2012A&A...547A.112M}.
However, to avoid losing transits within the \ariel{} observing window due to lack of precision on the ephemeris and/or on the orbital parameters,
it will be necessary to constrain the ephemeris
by taking advantage of ground-
(e.g. ExoClock Project,
TASTE, and NGTS) and
space-based telescopes (e.g. TESS, CHEOPS, and PLATO).
\par
We warn the reader that even if ArielRad takes into account different sources of stationary noise and
includes a noise margin, it cannot model time-correlated noise.
So, we can consider the values of the uncertainty on the $T_0$ obtained in our analysis as optimistic.
As soon as the on-flight performances of the satellite and of instruments will be precisely evaluated
we will be able to perform a deeper and more robust analysis of the impact of correlated noise
on the determination of the transit time and of its uncertainty.
A possible way to model correlated noise (or coloured noise, such as pink and red noise)
would be to use non-parametric models
such as ARMA or ARIMA processes, creating, for example,
a common coloured noise for both FGS1 and FGS2 and 
an additional coloured noise different for each filter.
We want to stress that currently the instrumental correlated noise is 
unknown and unpredictable and
also the stellar noise can vary star-by-star,
so it is not so trivial to properly model these noise components.
\par
 
In the future it would be interesting to repeat the analysis
taking into account multiple transits of the same planet
observed with \ariel{} (e.g. Tier 3 targets, about $\sim10\%$ of Tier 1) and 
assess the improvement on $\sigma_{T_0}$,
even if the main contribution derives from the sampling rate of the transit ingress and egress.
Furthermore, an MCMC analysis with a different algorithm,
such as Nested and Dynamic Nested Sampling \citep{Skilling2004AIPC..735..395S,skilling2006,Higson2019S&C....29..891H} 
implemented in \texttt{dynesty} \citep{Speagle2020MNRAS.493.3132S}
would help to sample in a more efficient and robust way a high dimensional space as in the modelling
of many transit light curves in different bands and modelling the dynamics of multi-planet systems.
\par

Furthermore, \ariel{} will be able for some orbital configurations to independently
detect TTV signals with about 10 transit observations,
allowing us to identify perturbing planets with masses spanning the Earth-Neptune range.
Combining RV data with transit times helps to put strong constraints on
bulk density and orbital parameters of the transiting planet
\citep[e.g. Kepler-19 and Kepler-9,][, respectively]{Malavolta2017AJ....153..224M, Borsato2019MNRAS.484.3233B}.
However, if the RV data have not enough precision to detect the external perturber, 
and taking into account the precision in the mass measurement of $~17\%$
for K2-24 c in \citet{Petigura2018AJ....156...89P},
we can say that ``in any case''
the detection of a TTV signal in \ariel{} data will allow us to determine
planetary masses of perturber with a precision better than $20\%$ in the Earth-Neptune regime.
\par

The \ariel{} target list is evolving with time and with new discoveries,
so at the moment is very difficult to give a quantitative estimates
of the number of known multi-planet system with TTV that \ariel{} will observe
and characterise.
We can expect about several tens of planets with significant TTV improvements,
and a similar number of targets with TTV,
whose contribution from \ariel{} will be important.
\citet{Hadden2019AJ....158..146H} predicted that about less than 4000 planet systems
will be discovered during the nominal mission of TESS
and about half of these will be in multi-planet systems.
30 of these multi-planet systems will have a measurable TTV signal and
only 10 will be characterised with precise mass measurement from TTV only.
Differently from missions like TESS or PLATO,
\ariel{} will point a single target, as CHEOPS,
and it will observe about 1000 targets,
roughly a quarter of the TESS predicted value.
So, we can assume for \ariel{} a similar number of the predicted TTV systems for TESS,
but only before launch we will know which planets 
will be observed and 
how many of these will be systems with TTV,
and it will mostly depend on the selection process and
on the schedulability of the targets.
\par

\begin{acknowledgements}
LBo, VNa, and GPi acknowledge the funding support from Italian Space Agency (ASI) regulated by ``Accordo ASI-INAF n. 2013-016-R.0 del 9 luglio 2013 e integrazione del 9 luglio 2015''. 
\end{acknowledgements}

%
%

\bibliographystyle{spbasic}      
\bibliography{biblio}   

\begin{thebibliography}{60}
\providecommand{\natexlab}[1]{#1}
\providecommand{\url}[1]{{#1}}
\providecommand{\urlprefix}{URL }
\expandafter\ifx\csname urlstyle\endcsname\relax
  \providecommand{\doi}[1]{DOI~\discretionary{}{}{}#1}\else
  \providecommand{\doi}{DOI~\discretionary{}{}{}\begingroup
  \urlstyle{rm}\Url}\fi
\providecommand{\eprint}[2][]{\url{#2}}

\bibitem[{{Agol} et~al.(2005){Agol}, {Steffen}, {Sari}, and
  {Clarkson}}]{Agol2005MNRAS.359..567A}
{Agol} E, {Steffen} J, {Sari} R, {Clarkson} W (2005) {On detecting terrestrial
  planets with timing of giant planet transits}. \mnras 359(2):567--579,
  \doi{10.1111/j.1365-2966.2005.08922.x}, \eprint{2005.08922}

\bibitem[{{Beichman} et~al.(2014){Beichman}, {Benneke}, {Knutson}, {Smith},
  {Lagage}, {Dressing}, {Latham}, {Lunine}, {Birkmann}, {Ferruit}, {Giardino},
  {Kempton}, {Carey}, {Krick}, {Deroo}, {Mand ell}, {Ressler}, {Shporer},
  {Swain}, {Vasisht}, {Ricker}, {Bouwman}, {Crossfield}, {Greene}, {Howell},
  {Christiansen}, {Ciardi}, {Clampin}, {Greenhouse}, {Sozzetti}, {Goudfrooij},
  {Hines}, {Keyes}, {Lee}, {McCullough}, {Robberto}, {Stansberry}, {Valenti},
  {Rieke}, {Rieke}, {Fortney}, {Bean}, {Kreidberg}, {Ehrenreich}, {Deming},
  {Albert}, {Doyon}, and {Sing}}]{Beichman2014PASP..126.1134B}
{Beichman} C, {Benneke} B, {Knutson} H, {Smith} R, {Lagage} PO, {Dressing} C,
  {Latham} D, {Lunine} J, {Birkmann} S, {Ferruit} P, {Giardino} G, {Kempton} E,
  {Carey} S, {Krick} J, {Deroo} PD, {Mand ell} A, {Ressler} ME, {Shporer} A,
  {Swain} M, {Vasisht} G, {Ricker} G, {Bouwman} J, {Crossfield} I, {Greene} T,
  {Howell} S, {Christiansen} J, {Ciardi} D, {Clampin} M, {Greenhouse} M,
  {Sozzetti} A, {Goudfrooij} P, {Hines} D, {Keyes} T, {Lee} J, {McCullough} P,
  {Robberto} M, {Stansberry} J, {Valenti} J, {Rieke} M, {Rieke} G, {Fortney} J,
  {Bean} J, {Kreidberg} L, {Ehrenreich} D, {Deming} D, {Albert} L, {Doyon} R,
  {Sing} D (2014) {Observations of Transiting Exoplanets with the James Webb
  Space Telescope (JWST)}. \pasp 126(946):1134, \doi{10.1086/679566}

\bibitem[{{Benz} et~al.(2020){Benz}, {Broeg}, {Fortier}, {Rando}, {Beck},
  {Beck}, {Queloz}, {Ehrenreich}, {Maxted}, {Isaak}, {Billot}, {Alibert},
  {Alonso}, {Ant{\'o}nio}, {Asquier}, {Bandy}, {B{\'a}rczy}, {Barrado},
  {Barros}, {Baumjohann}, {Bekkelien}, {Bergomi}, {Biondi}, {Bonfils},
  {Borsato}, {Brandeker}, {Busch}, {Cabrera}, {Cessa}, {Charnoz}, {Chazelas},
  {Collier Cameron}, {Corral Van Damme}, {Cortes}, {Davies}, {Deleuil},
  {Deline}, {Delrez}, {Demangeon}, {Demory}, {Erikson}, {Farinato}, {Fossati},
  {Fridlund}, {Futyan}, {Gandolfi}, {Garcia Munoz}, {Gillon}, {Guterman},
  {Gutierrez}, {Hasiba}, {Heng}, {Hernandez}, {Hoyer}, {Kiss}, {Kovacs},
  {Kuntzer}, {Laskar}, {Lecavelier des Etangs}, {Lendl}, {L{\'o}pez}, {Lora},
  {Lovis}, {L{\"u}ftinger}, {Magrin}, {Malvasio}, {Marafatto}, {Michaelis}, {de
  Miguel}, {Modrego}, {Munari}, {Nascimbeni}, {Olofsson}, {Ottacher},
  {Ottensamer}, {Pagano}, {Palacios}, {Pall{\'e}}, {Peter}, {Piazza}, {Piotto},
  {Pizarro}, {Pollaco}, {Ragazzoni}, {Ratti}, {Rauer}, {Ribas}, {Rieder},
  {Rohlfs}, {Safa}, {Salatti}, {Santos}, {Scandariato}, {S{\'e}gransan},
  {Simon}, {Smith}, {Sordet}, {Sousa}, {Steller}, {Szab{\'o}}, {Szoke},
  {Thomas}, {Tschentscher}, {Udry}, {Van Grootel}, {Viotto}, {Walter},
  {Walton}, {Wildi}, and {Wolter}}]{Benz2020ExA...tmp...53B}
{Benz} W, {Broeg} C, {Fortier} A, {Rando} N, {Beck} T, {Beck} M, {Queloz} D,
  {Ehrenreich} D, {Maxted} PFL, {Isaak} KG, {Billot} N, {Alibert} Y, {Alonso}
  R, {Ant{\'o}nio} C, {Asquier} J, {Bandy} T, {B{\'a}rczy} T, {Barrado} D,
  {Barros} SCC, {Baumjohann} W, {Bekkelien} A, {Bergomi} M, {Biondi} F,
  {Bonfils} X, {Borsato} L, {Brandeker} A, {Busch} MD, {Cabrera} J, {Cessa} V,
  {Charnoz} S, {Chazelas} B, {Collier Cameron} A, {Corral Van Damme} C,
  {Cortes} D, {Davies} MB, {Deleuil} M, {Deline} A, {Delrez} L, {Demangeon} O,
  {Demory} BO, {Erikson} A, {Farinato} J, {Fossati} L, {Fridlund} M, {Futyan}
  D, {Gandolfi} D, {Garcia Munoz} A, {Gillon} M, {Guterman} P, {Gutierrez} A,
  {Hasiba} J, {Heng} K, {Hernandez} E, {Hoyer} S, {Kiss} LL, {Kovacs} Z,
  {Kuntzer} T, {Laskar} J, {Lecavelier des Etangs} A, {Lendl} M, {L{\'o}pez} A,
  {Lora} I, {Lovis} C, {L{\"u}ftinger} T, {Magrin} D, {Malvasio} L, {Marafatto}
  L, {Michaelis} H, {de Miguel} D, {Modrego} D, {Munari} M, {Nascimbeni} V,
  {Olofsson} G, {Ottacher} H, {Ottensamer} R, {Pagano} I, {Palacios} R,
  {Pall{\'e}} E, {Peter} G, {Piazza} D, {Piotto} G, {Pizarro} A, {Pollaco} D,
  {Ragazzoni} R, {Ratti} F, {Rauer} H, {Ribas} I, {Rieder} M, {Rohlfs} R,
  {Safa} F, {Salatti} M, {Santos} NC, {Scandariato} G, {S{\'e}gransan} D,
  {Simon} AE, {Smith} AMS, {Sordet} M, {Sousa} SG, {Steller} M, {Szab{\'o}} GM,
  {Szoke} J, {Thomas} N, {Tschentscher} M, {Udry} S, {Van Grootel} V, {Viotto}
  V, {Walter} I, {Walton} NA, {Wildi} F, {Wolter} D (2020) {The CHEOPS
  mission}. Experimental Astronomy \doi{10.1007/s10686-020-09679-4},
  \eprint{2009.11633}

\bibitem[{{Borsato} et~al.(2014){Borsato}, {Marzari}, {Nascimbeni}, {Piotto},
  {Granata}, {Bedin}, and {Malavolta}}]{Borsato2014A&A...571A..38B}
{Borsato} L, {Marzari} F, {Nascimbeni} V, {Piotto} G, {Granata} V, {Bedin} LR,
  {Malavolta} L (2014) {TRADES: A new software to derive orbital parameters
  from observed transit times and radial velocities. Revisiting Kepler-11 and
  Kepler-9}. \aap 571:A38, \doi{10.1051/0004-6361/201424080},
  \eprint{1408.2844}

\bibitem[{{Borsato} et~al.(2019){Borsato}, {Malavolta}, {Piotto}, {Buchhave},
  {Mortier}, {Rice}, {Collier Cameron}, {Coffinet}, {Sozzetti}, {Charbonneau},
  {Cosentino}, {Dumusque}, {Figueira}, {Latham}, {Lopez-Morales}, {Mayor},
  {Micela}, {Molinari}, {Pepe}, {Phillips}, {Poretti}, {Udry}, and
  {Watson}}]{Borsato2019MNRAS.484.3233B}
{Borsato} L, {Malavolta} L, {Piotto} G, {Buchhave} LA, {Mortier} A, {Rice} K,
  {Collier Cameron} A, {Coffinet} A, {Sozzetti} A, {Charbonneau} D, {Cosentino}
  R, {Dumusque} X, {Figueira} P, {Latham} DW, {Lopez-Morales} M, {Mayor} M,
  {Micela} G, {Molinari} E, {Pepe} F, {Phillips} D, {Poretti} E, {Udry} S,
  {Watson} C (2019) {HARPS-N radial velocities confirm the low densities of the
  Kepler-9 planets}. \mnras 484(3):3233--3243, \doi{10.1093/mnras/stz181},
  \eprint{1901.05471}

\bibitem[{{Borucki} et~al.(2011){Borucki}, {Koch}, {Basri}, {Batalha}, {Brown},
  {Bryson}, {Caldwell}, {Christensen-Dalsgaard}, {Cochran}, {DeVore}, {Dunham},
  {Gautier}, {Geary}, {Gilliland}, {Gould}, {Howell}, {Jenkins}, {Latham},
  {Lissauer}, {Marcy}, {Rowe}, {Sasselov}, {Boss}, {Charbonneau}, {Ciardi},
  {Doyle}, {Dupree}, {Ford}, {Fortney}, {Holman}, {Seager}, {Steffen},
  {Tarter}, {Welsh}, {Allen}, {Buchhave}, {Christiansen}, {Clarke}, {Das},
  {D{\'e}sert}, {Endl}, {Fabrycky}, {Fressin}, {Haas}, {Horch}, {Howard},
  {Isaacson}, {Kjeldsen}, {Kolodziejczak}, {Kulesa}, {Li}, {Lucas}, {Machalek},
  {McCarthy}, {MacQueen}, {Meibom}, {Miquel}, {Prsa}, {Quinn}, {Quintana},
  {Ragozzine}, {Sherry}, {Shporer}, {Tenenbaum}, {Torres}, {Twicken}, {Van
  Cleve}, {Walkowicz}, {Witteborn}, and {Still}}]{Borucki2011ApJ...736...19B}
{Borucki} WJ, {Koch} DG, {Basri} G, {Batalha} N, {Brown} TM, {Bryson} ST,
  {Caldwell} D, {Christensen-Dalsgaard} J, {Cochran} WD, {DeVore} E, {Dunham}
  EW, {Gautier} I Thomas~N, {Geary} JC, {Gilliland} R, {Gould} A, {Howell} SB,
  {Jenkins} JM, {Latham} DW, {Lissauer} JJ, {Marcy} GW, {Rowe} J, {Sasselov} D,
  {Boss} A, {Charbonneau} D, {Ciardi} D, {Doyle} L, {Dupree} AK, {Ford} EB,
  {Fortney} J, {Holman} MJ, {Seager} S, {Steffen} JH, {Tarter} J, {Welsh} WF,
  {Allen} C, {Buchhave} LA, {Christiansen} JL, {Clarke} BD, {Das} S,
  {D{\'e}sert} JM, {Endl} M, {Fabrycky} D, {Fressin} F, {Haas} M, {Horch} E,
  {Howard} A, {Isaacson} H, {Kjeldsen} H, {Kolodziejczak} J, {Kulesa} C, {Li}
  J, {Lucas} PW, {Machalek} P, {McCarthy} D, {MacQueen} P, {Meibom} S, {Miquel}
  T, {Prsa} A, {Quinn} SN, {Quintana} EV, {Ragozzine} D, {Sherry} W, {Shporer}
  A, {Tenenbaum} P, {Torres} G, {Twicken} JD, {Van Cleve} J, {Walkowicz} L,
  {Witteborn} FC, {Still} M (2011) {Characteristics of Planetary Candidates
  Observed by Kepler. II. Analysis of the First Four Months of Data}. \apj
  736(1):19, \doi{10.1088/0004-637X/736/1/19}, \eprint{1102.0541}

\bibitem[{{Broeg} et~al.(2014){Broeg}, {Benz}, {Thomas}, and {Cheops
  Team}}]{Broeg2014CoSka..43..498B}
{Broeg} C, {Benz} W, {Thomas} N, {Cheops Team} (2014) {The CHEOPS mission}.
  Contributions of the Astronomical Observatory Skalnate Pleso 43(3):498--498

\bibitem[{{Delrez} et~al.(2018){Delrez}, {Gillon}, {Triaud}, {Demory}, {de
  Wit}, {Ingalls}, {Agol}, {Bolmont}, {Burdanov}, {Burgasser}, {Carey},
  {Jehin}, {Leconte}, {Lederer}, {Queloz}, {Selsis}, and {Van
  Grootel}}]{Delrez2018MNRAS.475.3577D}
{Delrez} L, {Gillon} M, {Triaud} AHMJ, {Demory} BO, {de Wit} J, {Ingalls} JG,
  {Agol} E, {Bolmont} E, {Burdanov} A, {Burgasser} AJ, {Carey} SJ, {Jehin} E,
  {Leconte} J, {Lederer} S, {Queloz} D, {Selsis} F, {Van Grootel} V (2018)
  {Early 2017 observations of TRAPPIST-1 with Spitzer}. \mnras
  475(3):3577--3597, \doi{10.1093/mnras/sty051}, \eprint{1801.02554}

\bibitem[{{Demory} et~al.(2016){Demory}, {Gillon}, {de Wit}, {Madhusudhan},
  {Bolmont}, {Heng}, {Kataria}, {Lewis}, {Hu}, {Krick}, {Stamenkovi{\'c}},
  {Benneke}, {Kane}, and {Queloz}}]{Demory2016Natur.532..207D}
{Demory} BO, {Gillon} M, {de Wit} J, {Madhusudhan} N, {Bolmont} E, {Heng} K,
  {Kataria} T, {Lewis} N, {Hu} R, {Krick} J, {Stamenkovi{\'c}} V, {Benneke} B,
  {Kane} S, {Queloz} D (2016) {A map of the large day-night temperature
  gradient of a super-Earth exoplanet}. \nat 532(7598):207--209,
  \doi{10.1038/nature17169}, \eprint{1604.05725}

\bibitem[{{Eastman} et~al.(2013){Eastman}, {Gaudi}, and
  {Agol}}]{Eastman2013PASP..125...83E}
{Eastman} J, {Gaudi} BS, {Agol} E (2013) {EXOFAST: A Fast Exoplanetary Fitting
  Suite in IDL}. \pasp 125(923):83, \doi{10.1086/669497}, \eprint{1206.5798}

\bibitem[{{Encrenaz} et~al.(2018){Encrenaz}, {Tinetti}, and
  {Coustenis}}]{Encrenaz2018ExA....46...31E}
{Encrenaz} T, {Tinetti} G, {Coustenis} A (2018) {Transit spectroscopy of
  temperate Jupiters with ARIEL: a feasibility study}. Experimental Astronomy
  46(1):31--44, \doi{10.1007/s10686-017-9561-2}

\bibitem[{{Espinoza} and {Jord{\'a}n}(2015)}]{Espinoza2015MNRAS.450.1879E}
{Espinoza} N, {Jord{\'a}n} A (2015) {Limb darkening and exoplanets: testing
  stellar model atmospheres and identifying biases in transit parameters}.
  \mnras 450(2):1879--1899, \doi{10.1093/mnras/stv744}, \eprint{1503.07020}

\bibitem[{{Foreman-Mackey} et~al.(2013){Foreman-Mackey}, {Conley}, {Meierjurgen
  Farr}, {Hogg}, {Lang}, {Marshall}, {Price-Whelan}, {Sanders}, and
  {Zuntz}}]{DFM2013ascl.soft03002F}
{Foreman-Mackey} D, {Conley} A, {Meierjurgen Farr} W, {Hogg} DW, {Lang} D,
  {Marshall} P, {Price-Whelan} A, {Sanders} J, {Zuntz} J (2013) {emcee: The
  MCMC Hammer}. \eprint{1303.002}

\bibitem[{{Foreman-Mackey} et~al.(2019){Foreman-Mackey}, {Farr}, {Sinha},
  {Archibald}, {Hogg}, {Sanders}, {Zuntz}, {Williams}, {Nelson}, {de
  Val-Borro}, {Erhardt}, {Pashchenko}, and {Pla}}]{DFM2019JOSS....4.1864F}
{Foreman-Mackey} D, {Farr} W, {Sinha} M, {Archibald} A, {Hogg} D, {Sanders} J,
  {Zuntz} J, {Williams} P, {Nelson} A, {de Val-Borro} M, {Erhardt} T,
  {Pashchenko} I, {Pla} O (2019) {emcee v3: A Python ensemble sampling toolkit
  for affine-invariant MCMC}. The Journal of Open Source Software 4(43):1864,
  \doi{10.21105/joss.01864}, \eprint{1911.07688}

\bibitem[{{Goodman} and {Weare}(2010)}]{GoodmanWeare2010CAMCS...5...65G}
{Goodman} J, {Weare} J (2010) {Ensemble samplers with affine invariance}.
  Communications in Applied Mathematics and Computational Science 5(1):65--80,
  \doi{10.2140/camcos.2010.5.65}

\bibitem[{{Granata} et~al.(2014){Granata}, {Nascimbeni}, {Piotto}, {Bedin},
  {Borsato}, {Cunial}, {Damasso}, and {Malavolta
  }}]{Granata2014AN....335..797G}
{Granata} V, {Nascimbeni} V, {Piotto} G, {Bedin} LR, {Borsato} L, {Cunial} A,
  {Damasso} M, {Malavolta } L (2014) {TASTE IV: Refining ephemeris and orbital
  parameters for HAT-P-20b and WASP-1b}. Astronomische Nachrichten 335(8):797,
  \doi{10.1002/asna.201412072}, \eprint{1405.3288}

\bibitem[{{Greene} et~al.(2017){Greene}, {Schlawin}, {Beichman}, {Line},
  {Fortney}, {Fraine}, and {JWST NIRCam Team}}]{Greene2017AAS...23030901G}
{Greene} TP, {Schlawin} E, {Beichman} CA, {Line} MR, {Fortney} JJ, {Fraine} J,
  {JWST NIRCam Team} (2017) {Characterizing transiting exoplanet atmospheres
  using NIRCam grism spectra}. In: American Astronomical Society Meeting
  Abstracts \#230, American Astronomical Society Meeting Abstracts, vol 230, p
  309.01

\bibitem[{{Hadden} et~al.(2019){Hadden}, {Barclay}, {Payne}, and
  {Holman}}]{Hadden2019AJ....158..146H}
{Hadden} S, {Barclay} T, {Payne} MJ, {Holman} MJ (2019) {Prospects for TTV
  Detection and Dynamical Constraints with TESS}. \aj 158(4):146,
  \doi{10.3847/1538-3881/ab384c}, \eprint{1811.01970}

\bibitem[{{Higson} et~al.(2019){Higson}, {Handley}, {Hobson}, and
  {Lasenby}}]{Higson2019S&C....29..891H}
{Higson} E, {Handley} W, {Hobson} M, {Lasenby} A (2019) {Dynamic nested
  sampling: an improved algorithm for parameter estimation and evidence
  calculation}. Statistics and Computing 29(5):891--913,
  \doi{10.1007/s11222-018-9844-0}, \eprint{1704.03459}

\bibitem[{{Holczer} et~al.(2016){Holczer}, {Mazeh}, {Nachmani},
  {Jontof-Hutter}, {Ford}, {Fabrycky}, {Ragozzine}, {Kane}, and
  {Steffen}}]{Holczer2016ApJS..225....9H}
{Holczer} T, {Mazeh} T, {Nachmani} G, {Jontof-Hutter} D, {Ford} EB, {Fabrycky}
  D, {Ragozzine} D, {Kane} M, {Steffen} JH (2016) {Transit Timing Observations
  from Kepler. IX. Catalog of the Full Long-cadence Data Set}. \apjs 225(1):9,
  \doi{10.3847/0067-0049/225/1/9}, \eprint{1606.01744}

\bibitem[{{Holman} and {Murray}(2005)}]{HolmanMurray2005Sci...307.1288H}
{Holman} MJ, {Murray} NW (2005) {The Use of Transit Timing to Detect
  Terrestrial-Mass Extrasolar Planets}. Science 307(5713):1288--1291,
  \doi{10.1126/science.1107822}, \eprint{astro-ph/0412028}

\bibitem[{{Holman} et~al.(2010){Holman}, {Fabrycky}, {Ragozzine}, {Ford},
  {Steffen}, {Welsh}, {Lissauer}, {Latham}, {Marcy}, {Walkowicz}, {Batalha},
  {Jenkins}, {Rowe}, {Cochran}, {Fressin}, {Torres}, {Buchhave}, {Sasselov},
  {Borucki}, {Koch}, {Basri}, {Brown}, {Caldwell}, {Charbonneau}, {Dunham},
  {Gautier}, {Geary}, {Gilliland}, {Haas}, {Howell}, {Ciardi}, {Endl},
  {Fischer}, {F{\"u}r{\'e}sz}, {Hartman}, {Isaacson}, {Johnson}, {MacQueen},
  {Moorhead}, {Morehead}, and {Orosz}}]{Holman2010Sci...330...51H}
{Holman} MJ, {Fabrycky} DC, {Ragozzine} D, {Ford} EB, {Steffen} JH, {Welsh} WF,
  {Lissauer} JJ, {Latham} DW, {Marcy} GW, {Walkowicz} LM, {Batalha} NM,
  {Jenkins} JM, {Rowe} JF, {Cochran} WD, {Fressin} F, {Torres} G, {Buchhave}
  LA, {Sasselov} DD, {Borucki} WJ, {Koch} DG, {Basri} G, {Brown} TM, {Caldwell}
  DA, {Charbonneau} D, {Dunham} EW, {Gautier} TN, {Geary} JC, {Gilliland} RL,
  {Haas} MR, {Howell} SB, {Ciardi} DR, {Endl} M, {Fischer} D, {F{\"u}r{\'e}sz}
  G, {Hartman} JD, {Isaacson} H, {Johnson} JA, {MacQueen} PJ, {Moorhead} AV,
  {Morehead} RC, {Orosz} JA (2010) {Kepler-9: A System of Multiple Planets
  Transiting a Sun-Like Star, Confirmed by Timing Variations}. Science
  330(6000):51, \doi{10.1126/science.1195778}

\bibitem[{{Howell} et~al.(2014){Howell}, {Sobeck}, {Haas}, {Still}, {Barclay},
  {Mullally}, {Troeltzsch}, {Aigrain}, {Bryson}, {Caldwell}, {Chaplin},
  {Cochran}, {Huber}, {Marcy}, {Miglio}, {Najita}, {Smith}, {Twicken}, and
  {Fortney}}]{Howell2014PASP..126..398H}
{Howell} SB, {Sobeck} C, {Haas} M, {Still} M, {Barclay} T, {Mullally} F,
  {Troeltzsch} J, {Aigrain} S, {Bryson} ST, {Caldwell} D, {Chaplin} WJ,
  {Cochran} WD, {Huber} D, {Marcy} GW, {Miglio} A, {Najita} JR, {Smith} M,
  {Twicken} JD, {Fortney} JJ (2014) {The K2 Mission: Characterization and Early
  Results}. \pasp 126(938):398, \doi{10.1086/676406}, \eprint{1402.5163}

\bibitem[{{Husser} et~al.(2013){Husser}, {Wende-von Berg}, {Dreizler},
  {Homeier}, {Reiners}, {Barman}, and {Hauschildt}}]{Husser2013A&A...553A...6H}
{Husser} TO, {Wende-von Berg} S, {Dreizler} S, {Homeier} D, {Reiners} A,
  {Barman} T, {Hauschildt} PH (2013) {A new extensive library of PHOENIX
  stellar atmospheres and synthetic spectra}. \aap 553:A6,
  \doi{10.1051/0004-6361/201219058}, \eprint{1303.5632}

\bibitem[{{Kipping}(2010{\natexlab{a}})}]{Kipping2010MNRAS.408.1758K}
{Kipping} DM (2010{\natexlab{a}}) {Binning is sinning: morphological
  light-curve distortions due to finite integration time}. \mnras
  408(3):1758--1769, \doi{10.1111/j.1365-2966.2010.17242.x}, \eprint{1004.3741}

\bibitem[{{Kipping}(2010{\natexlab{b}})}]{Kipping2010MNRAS.407..301K}
{Kipping} DM (2010{\natexlab{b}}) {Investigations of approximate expressions
  for the transit duration}. \mnras 407(1):301--313,
  \doi{10.1111/j.1365-2966.2010.16894.x}, \eprint{1004.3819}

\bibitem[{{Kipping}(2013)}]{Kipping2013MNRAS.435.2152K}
{Kipping} DM (2013) {Efficient, uninformative sampling of limb darkening
  coefficients for two-parameter laws}. \mnras 435(3):2152--2160,
  \doi{10.1093/mnras/stt1435}, \eprint{1308.0009}

\bibitem[{{Kreidberg}(2015)}]{Kreidberg2015PASP..127.1161K}
{Kreidberg} L (2015) {batman: BAsic Transit Model cAlculatioN in Python}. \pasp
  127(957):1161, \doi{10.1086/683602}, \eprint{1507.08285}

\bibitem[{{Kurucz}(1979)}]{Kurucz1979ApJS...40....1K}
{Kurucz} RL (1979) {Model atmospheres for G, F, A, B, and O stars.} \apjs
  40:1--340, \doi{10.1086/190589}

\bibitem[{{Lissauer} et~al.(2011){Lissauer}, {Fabrycky}, {Ford}, {Borucki},
  {Fressin}, {Marcy}, {Orosz}, {Rowe}, {Torres}, {Welsh}, {Batalha}, {Bryson},
  {Buchhave}, {Caldwell}, {Carter}, {Charbonneau}, {Christiansen}, {Cochran},
  {Desert}, {Dunham}, {Fanelli}, {Fortney}, {Gautier}, {Geary}, {Gilliland},
  {Haas}, {Hall}, {Holman}, {Koch}, {Latham}, {Lopez}, {McCauliff}, {Miller},
  {Morehead}, {Quintana}, {Ragozzine}, {Sasselov}, {Short}, and
  {Steffen}}]{Lissauer2011Natur.470...53L}
{Lissauer} JJ, {Fabrycky} DC, {Ford} EB, {Borucki} WJ, {Fressin} F, {Marcy} GW,
  {Orosz} JA, {Rowe} JF, {Torres} G, {Welsh} WF, {Batalha} NM, {Bryson} ST,
  {Buchhave} LA, {Caldwell} DA, {Carter} JA, {Charbonneau} D, {Christiansen}
  JL, {Cochran} WD, {Desert} JM, {Dunham} EW, {Fanelli} MN, {Fortney} JJ,
  {Gautier} I Thomas~N, {Geary} JC, {Gilliland} RL, {Haas} MR, {Hall} JR,
  {Holman} MJ, {Koch} DG, {Latham} DW, {Lopez} E, {McCauliff} S, {Miller} N,
  {Morehead} RC, {Quintana} EV, {Ragozzine} D, {Sasselov} D, {Short} DR,
  {Steffen} JH (2011) {A closely packed system of low-mass, low-density planets
  transiting Kepler-11}. \nat 470(7332):53--58, \doi{10.1038/nature09760},
  \eprint{1102.0291}

\bibitem[{{Lissauer} et~al.(2013){Lissauer}, {Jontof-Hutter}, {Rowe},
  {Fabrycky}, {Lopez}, {Agol}, {Marcy}, {Deck}, {Fischer}, {Fortney}, {Howell},
  {Isaacson}, {Jenkins}, {Kolbl}, {Sasselov}, {Short}, and
  {Welsh}}]{Lissauer2013ApJ...770..131L}
{Lissauer} JJ, {Jontof-Hutter} D, {Rowe} JF, {Fabrycky} DC, {Lopez} ED, {Agol}
  E, {Marcy} GW, {Deck} KM, {Fischer} DA, {Fortney} JJ, {Howell} SB, {Isaacson}
  H, {Jenkins} JM, {Kolbl} R, {Sasselov} D, {Short} DR, {Welsh} WF (2013) {All
  Six Planets Known to Orbit Kepler-11 Have Low Densities}. \apj 770(2):131,
  \doi{10.1088/0004-637X/770/2/131}, \eprint{1303.0227}

\bibitem[{{Malavolta} et~al.(2017){Malavolta}, {Borsato}, {Granata}, {Piotto},
  {Lopez}, {Vanderburg}, {Figueira}, {Mortier}, {Nascimbeni}, {Affer},
  {Bonomo}, {Bouchy}, {Buchhave}, {Charbonneau}, {Collier Cameron},
  {Cosentino}, {Dressing}, {Dumusque}, {Fiorenzano}, {Harutyunyan}, {Haywood},
  {Johnson}, {Latham}, {Lopez-Morales}, {Lovis}, {Mayor}, {Micela}, {Molinari},
  {Motalebi}, {Pepe}, {Phillips}, {Pollacco}, {Queloz}, {Rice}, {Sasselov},
  {S{\'e}gransan}, {Sozzetti}, {Udry}, and
  {Watson}}]{Malavolta2017AJ....153..224M}
{Malavolta} L, {Borsato} L, {Granata} V, {Piotto} G, {Lopez} E, {Vanderburg} A,
  {Figueira} P, {Mortier} A, {Nascimbeni} V, {Affer} L, {Bonomo} AS, {Bouchy}
  F, {Buchhave} LA, {Charbonneau} D, {Collier Cameron} A, {Cosentino} R,
  {Dressing} CD, {Dumusque} X, {Fiorenzano} AFM, {Harutyunyan} A, {Haywood} RD,
  {Johnson} JA, {Latham} DW, {Lopez-Morales} M, {Lovis} C, {Mayor} M, {Micela}
  G, {Molinari} E, {Motalebi} F, {Pepe} F, {Phillips} DF, {Pollacco} D,
  {Queloz} D, {Rice} K, {Sasselov} D, {S{\'e}gransan} D, {Sozzetti} A, {Udry}
  S, {Watson} C (2017) {The Kepler-19 System: A Thick-envelope Super-Earth with
  Two Neptune-mass Companions Characterized Using Radial Velocities and Transit
  Timing Variations}. \aj 153(5):224, \doi{10.3847/1538-3881/aa6897},
  \eprint{1703.06885}

\bibitem[{{Mazeh} et~al.(2013){Mazeh}, {Nachmani}, {Holczer}, {Fabrycky},
  {Ford}, {Sanchis-Ojeda}, {Sokol}, {Rowe}, {Zucker}, {Agol}, {Carter},
  {Lissauer}, {Quintana}, {Ragozzine}, {Steffen}, and
  {Welsh}}]{Mazeh2013ApJS..208...16M}
{Mazeh} T, {Nachmani} G, {Holczer} T, {Fabrycky} DC, {Ford} EB, {Sanchis-Ojeda}
  R, {Sokol} G, {Rowe} JF, {Zucker} S, {Agol} E, {Carter} JA, {Lissauer} JJ,
  {Quintana} EV, {Ragozzine} D, {Steffen} JH, {Welsh} W (2013) {Transit Timing
  Observations from Kepler. VIII. Catalog of Transit Timing Measurements of the
  First Twelve Quarters}. \apjs 208(2):16, \doi{10.1088/0067-0049/208/2/16},
  \eprint{1301.5499}

\bibitem[{{Miralda-Escud{\'e}}(2002)}]{Miralda-Escude2002ApJ...564.1019M}
{Miralda-Escud{\'e}} J (2002) {Orbital Perturbations of Transiting Planets: A
  Possible Method to Measure Stellar Quadrupoles and to Detect Earth-Mass
  Planets}. \apj 564(2):1019--1023, \doi{10.1086/324279},
  \eprint{astro-ph/0104034}

\bibitem[{{Molli{\`e}re} et~al.(2017){Molli{\`e}re}, {van Boekel}, {Bouwman},
  {Henning}, {Lagage}, and {Min}}]{Molliere2017A&A...600A..10M}
{Molli{\`e}re} P, {van Boekel} R, {Bouwman} J, {Henning} T, {Lagage} PO, {Min}
  M (2017) {Observing transiting planets with JWST. Prime targets and their
  synthetic spectral observations}. \aap 600:A10,
  \doi{10.1051/0004-6361/201629800}, \eprint{1611.08608}

\bibitem[{{Mordasini} et~al.(2012){Mordasini}, {Alibert}, {Georgy},
  {Dittkrist}, {Klahr}, and {Henning}}]{Mordasini2012A&A...547A.112M}
{Mordasini} C, {Alibert} Y, {Georgy} C, {Dittkrist} KM, {Klahr} H, {Henning} T
  (2012) {Characterization of exoplanets from their formation. II. The
  planetary mass-radius relationship}. \aap 547:A112,
  \doi{10.1051/0004-6361/201118464}, \eprint{1206.3303}

\bibitem[{{Mugnai} et~al.(2020){Mugnai}, {Pascale}, {Edwards}, {Papageorgiou},
  and {Sarkar}}]{Mugnai2020ExA....50..303M}
{Mugnai} LV, {Pascale} E, {Edwards} B, {Papageorgiou} A, {Sarkar} S (2020)
  {ArielRad: the Ariel radiometric model}. Experimental Astronomy
  50(2-3):303--328, \doi{10.1007/s10686-020-09676-7}, \eprint{2009.07824}

\bibitem[{{Nascimbeni} et~al.(2011){Nascimbeni}, {Piotto}, {Bedin}, and
  {Damasso}}]{Nascimbeni2011A&A...527A..85N}
{Nascimbeni} V, {Piotto} G, {Bedin} LR, {Damasso} M (2011) {TASTE: The Asiago
  Search for Transit timing variations of Exoplanets. I. Overview and improved
  parameters for HAT-P-3b and HAT-P-14b}. \aap 527:A85,
  \doi{10.1051/0004-6361/201015199}, \eprint{1011.6395}

\bibitem[{{Nesvorn{\'y}} et~al.(2013){Nesvorn{\'y}}, {Kipping}, {Terrell},
  {Hartman}, {Bakos}, and {Buchhave}}]{Nesvorny2013ApJ...777....3N}
{Nesvorn{\'y}} D, {Kipping} D, {Terrell} D, {Hartman} J, {Bakos} G{\'A},
  {Buchhave} LA (2013) {KOI-142, The King of Transit Variations, is a Pair of
  Planets near the 2:1 Resonance}. \apj 777(1):3,
  \doi{10.1088/0004-637X/777/1/3}, \eprint{1304.4283}

\bibitem[{{Pascale} et~al.(2018){Pascale}, {Bezawada}, {Barstow}, {Beaulieu},
  {Bowles}, {Coud{\'e} du Foresto}, {Coustenis}, {Decin}, {Drossart},
  {Eccleston}, {Encrenaz}, {Forget}, {Griffin}, {G{\"u}del}, {Hartogh},
  {Heske}, {Lagage}, {Leconte}, {Malaguti}, {Micela}, {Middleton}, {Min},
  {Moneti}, {Morales}, {Mugnai}, {Ollivier}, {Pace}, {Papageorgiou},
  {Pilbratt}, {Puig}, {Rataj}, {Ray}, {Ribas}, {Rocchetto}, {Sarkar}, {Selsis},
  {Taylor}, {Tennyson}, {Tinetti}, {Turrini}, {Vandenbussche}, {Venot},
  {Waldmann}, {Wolkenberg}, {Wright}, {Zapatero Osorio}, and
  {Zingales}}]{Pascale2018SPIE10698E..0HP}
{Pascale} E, {Bezawada} N, {Barstow} J, {Beaulieu} JP, {Bowles} N, {Coud{\'e}
  du Foresto} V, {Coustenis} A, {Decin} L, {Drossart} P, {Eccleston} P,
  {Encrenaz} T, {Forget} F, {Griffin} M, {G{\"u}del} M, {Hartogh} P, {Heske} A,
  {Lagage} PO, {Leconte} J, {Malaguti} P, {Micela} G, {Middleton} K, {Min} M,
  {Moneti} A, {Morales} JC, {Mugnai} L, {Ollivier} M, {Pace} E, {Papageorgiou}
  A, {Pilbratt} G, {Puig} L, {Rataj} M, {Ray} T, {Ribas} I, {Rocchetto} M,
  {Sarkar} S, {Selsis} F, {Taylor} W, {Tennyson} J, {Tinetti} G, {Turrini} D,
  {Vandenbussche} B, {Venot} O, {Waldmann} IP, {Wolkenberg} P, {Wright} G,
  {Zapatero Osorio} MR, {Zingales} T (2018) {The ARIEL space mission}. In:
  \procspie, Society of Photo-Optical Instrumentation Engineers (SPIE)
  Conference Series, vol 10698, p 106980H, \doi{10.1117/12.2311838}

\bibitem[{{Perryman}(2018)}]{Perryman2018exha.book.....P}
{Perryman} M (2018) {The Exoplanet Handbook}

\bibitem[{{Petigura} et~al.(2016){Petigura}, {Howard}, {Lopez}, {Deck},
  {Fulton}, {Crossfield}, {Ciardi}, {Chiang}, {Lee}, {Isaacson}, {Beichman},
  {Hansen}, {Schlieder}, and {Sinukoff}}]{Petigura2016ApJ...818...36P}
{Petigura} EA, {Howard} AW, {Lopez} ED, {Deck} KM, {Fulton} BJ, {Crossfield}
  IJM, {Ciardi} DR, {Chiang} E, {Lee} EJ, {Isaacson} H, {Beichman} CA, {Hansen}
  BMS, {Schlieder} JE, {Sinukoff} E (2016) {Two Transiting Low Density
  Sub-Saturns from K2}. \apj 818(1):36, \doi{10.3847/0004-637X/818/1/36},
  \eprint{1511.04497}

\bibitem[{{Petigura} et~al.(2018){Petigura}, {Benneke}, {Batygin}, {Fulton},
  {Werner}, {Krick}, {Gorjian}, {Sinukoff}, {Deck}, {Mills}, and
  {Deming}}]{Petigura2018AJ....156...89P}
{Petigura} EA, {Benneke} B, {Batygin} K, {Fulton} BJ, {Werner} M, {Krick} JE,
  {Gorjian} V, {Sinukoff} E, {Deck} KM, {Mills} SM, {Deming} D (2018) {Dynamics
  and Formation of the Near-resonant K2-24 System: Insights from Transit-timing
  Variations and Radial Velocities}. \aj 156(3):89,
  \doi{10.3847/1538-3881/aaceac}, \eprint{1806.08959}

\bibitem[{{Pilbratt}(2019)}]{Pilbratt2019ESS.....450304P}
{Pilbratt} G (2019) {ARIEL: ESA's Mission to Study the Nature of Exoplanets}.
  In: AAS/Division for Extreme Solar Systems Abstracts, AAS/Division for
  Extreme Solar Systems Abstracts, vol~51, p 503.04

\bibitem[{{Price} and {Rogers}(2014)}]{PriceRogers2014ApJ...794...92P}
{Price} EM, {Rogers} LA (2014) {Transit Light Curves with Finite Integration
  Time: Fisher Information Analysis}. \apj 794(1):92,
  \doi{10.1088/0004-637X/794/1/92}, \eprint{1408.4124}

\bibitem[{{Puig} et~al.(2018){Puig}, {Pilbratt}, {Heske}, {Escudero},
  {Crouzet}, {de Vogeleer}, {Symonds}, {Kohley}, {Drossart}, {Eccleston},
  {Hartogh}, {Leconte}, {Micela}, {Ollivier}, {Tinetti}, {Turrini},
  {Vandenbussche}, and {Wolkenberg}}]{Puig2018ExA....46..211P}
{Puig} L, {Pilbratt} G, {Heske} A, {Escudero} I, {Crouzet} PE, {de Vogeleer} B,
  {Symonds} K, {Kohley} R, {Drossart} P, {Eccleston} P, {Hartogh} P, {Leconte}
  J, {Micela} G, {Ollivier} M, {Tinetti} G, {Turrini} D, {Vandenbussche} B,
  {Wolkenberg} P (2018) {The Phase A study of the ESA M4 mission candidate
  ARIEL}. Experimental Astronomy 46(1):211--239,
  \doi{10.1007/s10686-018-9604-3}

\bibitem[{{Rauer} et~al.(2014){Rauer}, {Catala}, {Aerts}, {Appourchaux},
  {Benz}, {Brandeker}, {Christensen-Dalsgaard}, {Deleuil}, {Gizon}, {Goupil},
  {G{\"u}del}, {Janot-Pacheco}, {Mas-Hesse}, {Pagano}, {Piotto}, {Pollacco},
  {Santos}, {Smith}, {Su\'{a}rez}, {Szab{\'o}}, {Udry}, {Adibekyan}, {Alibert},
  {Almenara}, {Amaro-Seoane}, {Eiff}, {Asplund}, {Antonello}, {Barnes},
  {Baudin}, {Belkacem}, {Bergemann}, {Bihain}, {Birch}, {Bonfils}, {Boisse},
  {Bonomo}, {Borsa}, {Brand {\~a}o}, {Brocato}, {Brun}, {Burleigh}, {Burston},
  {Cabrera}, {Cassisi}, {Chaplin}, {Charpinet}, {Chiappini}, {Church},
  {Csizmadia}, {Cunha}, {Damasso}, {Davies}, {Deeg}, {D{\'\i}az}, {Dreizler},
  {Dreyer}, {Eggenberger}, {Ehrenreich}, {Eigm{\"u}ller}, {Erikson}, {Farmer},
  {Feltzing}, {de Oliveira Fialho}, {Figueira}, {Forveille}, {Fridlund},
  {Garc{\'\i}a}, {Giommi}, {Giuffrida}, {Godolt}, {Gomes da Silva}, {Granzer},
  {Grenfell}, {Grotsch-Noels}, {G{\"u}nther}, {Haswell}, {Hatzes},
  {H{\'e}brard}, {Hekker}, {Helled}, {Heng}, {Jenkins}, {Johansen},
  {Khodachenko}, {Kislyakova}, {Kley}, {Kolb}, {Krivova}, {Kupka}, {Lammer},
  {Lanza}, {Lebreton}, {Magrin}, {Marcos-Arenal}, {Marrese}, {Marques},
  {Martins}, {Mathis}, {Mathur}, {Messina}, {Miglio}, {Montalban}, {Montalto},
  {Monteiro}, {Moradi}, {Moravveji}, {Mordasini}, {Morel}, {Mortier},
  {Nascimbeni}, {Nelson}, {Nielsen}, {Noack}, {Norton}, {Ofir}, {Oshagh},
  {Ouazzani}, {P{\'a}pics}, {Parro}, {Petit}, {Plez}, {Poretti}, {Quirrenbach},
  {Ragazzoni}, {Raimondo}, {Rainer}, {Reese}, {Redmer}, {Reffert},
  {Rojas-Ayala}, {Roxburgh}, {Salmon}, {Santerne}, {Schneider}, {Schou},
  {Schuh}, {Schunker}, {Silva-Valio}, {Silvotti}, {Skillen}, {Snellen}, {Sohl},
  {Sousa}, {Sozzetti}, {Stello}, {Strassmeier}, {{\v{S}}vanda}, {Szab{\'o}},
  {Tkachenko}, {Valencia}, {Van Grootel}, {Vauclair}, {Ventura}, {Wagner},
  {Walton}, {Weingrill}, {Werner}, {Wheatley}, and
  {Zwintz}}]{Rauer2014ExA....38..249R}
{Rauer} H, {Catala} C, {Aerts} C, {Appourchaux} T, {Benz} W, {Brandeker} A,
  {Christensen-Dalsgaard} J, {Deleuil} M, {Gizon} L, {Goupil} MJ, {G{\"u}del}
  M, {Janot-Pacheco} E, {Mas-Hesse} M, {Pagano} I, {Piotto} G, {Pollacco} D,
  {Santos} C, {Smith} A, {Su\'{a}rez} JC, {Szab{\'o}} R, {Udry} S, {Adibekyan}
  V, {Alibert} Y, {Almenara} JM, {Amaro-Seoane} P, {Eiff} MAv, {Asplund} M,
  {Antonello} E, {Barnes} S, {Baudin} F, {Belkacem} K, {Bergemann} M, {Bihain}
  G, {Birch} AC, {Bonfils} X, {Boisse} I, {Bonomo} AS, {Borsa} F, {Brand
  {\~a}o} IM, {Brocato} E, {Brun} S, {Burleigh} M, {Burston} R, {Cabrera} J,
  {Cassisi} S, {Chaplin} W, {Charpinet} S, {Chiappini} C, {Church} RP,
  {Csizmadia} S, {Cunha} M, {Damasso} M, {Davies} MB, {Deeg} HJ, {D{\'\i}az}
  RF, {Dreizler} S, {Dreyer} C, {Eggenberger} P, {Ehrenreich} D,
  {Eigm{\"u}ller} P, {Erikson} A, {Farmer} R, {Feltzing} S, {de Oliveira
  Fialho} F, {Figueira} P, {Forveille} T, {Fridlund} M, {Garc{\'\i}a} RA,
  {Giommi} P, {Giuffrida} G, {Godolt} M, {Gomes da Silva} J, {Granzer} T,
  {Grenfell} JL, {Grotsch-Noels} A, {G{\"u}nther} E, {Haswell} CA, {Hatzes} AP,
  {H{\'e}brard} G, {Hekker} S, {Helled} R, {Heng} K, {Jenkins} JM, {Johansen}
  A, {Khodachenko} ML, {Kislyakova} KG, {Kley} W, {Kolb} U, {Krivova} N,
  {Kupka} F, {Lammer} H, {Lanza} AF, {Lebreton} Y, {Magrin} D, {Marcos-Arenal}
  P, {Marrese} PM, {Marques} JP, {Martins} J, {Mathis} S, {Mathur} S, {Messina}
  S, {Miglio} A, {Montalban} J, {Montalto} M, {Monteiro} MJPFG, {Moradi} H,
  {Moravveji} E, {Mordasini} C, {Morel} T, {Mortier} A, {Nascimbeni} V,
  {Nelson} RP, {Nielsen} MB, {Noack} L, {Norton} AJ, {Ofir} A, {Oshagh} M,
  {Ouazzani} RM, {P{\'a}pics} P, {Parro} VC, {Petit} P, {Plez} B, {Poretti} E,
  {Quirrenbach} A, {Ragazzoni} R, {Raimondo} G, {Rainer} M, {Reese} DR,
  {Redmer} R, {Reffert} S, {Rojas-Ayala} B, {Roxburgh} IW, {Salmon} S,
  {Santerne} A, {Schneider} J, {Schou} J, {Schuh} S, {Schunker} H,
  {Silva-Valio} A, {Silvotti} R, {Skillen} I, {Snellen} I, {Sohl} F, {Sousa}
  SG, {Sozzetti} A, {Stello} D, {Strassmeier} KG, {{\v{S}}vanda} M, {Szab{\'o}}
  GM, {Tkachenko} A, {Valencia} D, {Van Grootel} V, {Vauclair} SD, {Ventura} P,
  {Wagner} FW, {Walton} NA, {Weingrill} J, {Werner} SC, {Wheatley} PJ, {Zwintz}
  K (2014) {The PLATO 2.0 mission}. Experimental Astronomy 38(1-2):249--330,
  \doi{10.1007/s10686-014-9383-4}, \eprint{1310.0696}

\bibitem[{{Ricker} et~al.(2014){Ricker}, {Winn}, {Vanderspek}, {Latham},
  {Bakos}, {Bean}, {Berta-Thompson}, {Brown}, {Buchhave}, {Butler}, {Butler},
  {Chaplin}, {Charbonneau}, {Christensen-Dalsgaard}, {Clampin}, {Deming},
  {Doty}, {De Lee}, {Dressing}, {Dunham}, {Endl}, {Fressin}, {Ge}, {Henning},
  {Holman}, {Howard}, {Ida}, {Jenkins}, {Jernigan}, {Johnson}, {Kaltenegger},
  {Kawai}, {Kjeldsen}, {Laughlin}, {Levine}, {Lin}, {Lissauer}, {MacQueen},
  {Marcy}, {McCullough}, {Morton}, {Narita}, {Paegert}, {Palle}, {Pepe},
  {Pepper}, {Quirrenbach}, {Rinehart}, {Sasselov}, {Sato}, {Seager},
  {Sozzetti}, {Stassun}, {Sullivan}, {Szentgyorgyi}, {Torres}, {Udry}, and
  {Villasenor}}]{Ricker2014SPIE.9143E..20R}
{Ricker} GR, {Winn} JN, {Vanderspek} R, {Latham} DW, {Bakos} G{\'A}, {Bean} JL,
  {Berta-Thompson} ZK, {Brown} TM, {Buchhave} L, {Butler} NR, {Butler} RP,
  {Chaplin} WJ, {Charbonneau} D, {Christensen-Dalsgaard} J, {Clampin} M,
  {Deming} D, {Doty} J, {De Lee} N, {Dressing} C, {Dunham} EW, {Endl} M,
  {Fressin} F, {Ge} J, {Henning} T, {Holman} MJ, {Howard} AW, {Ida} S,
  {Jenkins} J, {Jernigan} G, {Johnson} JA, {Kaltenegger} L, {Kawai} N,
  {Kjeldsen} H, {Laughlin} G, {Levine} AM, {Lin} D, {Lissauer} JJ, {MacQueen}
  P, {Marcy} G, {McCullough} PR, {Morton} TD, {Narita} N, {Paegert} M, {Palle}
  E, {Pepe} F, {Pepper} J, {Quirrenbach} A, {Rinehart} SA, {Sasselov} D, {Sato}
  B, {Seager} S, {Sozzetti} A, {Stassun} KG, {Sullivan} P, {Szentgyorgyi} A,
  {Torres} G, {Udry} S, {Villasenor} J (2014) {Transiting Exoplanet Survey
  Satellite (TESS)}. In: \procspie, Society of Photo-Optical Instrumentation
  Engineers (SPIE) Conference Series, vol 9143, p 914320,
  \doi{10.1117/12.2063489}, \eprint{1406.0151}

\bibitem[{{Skilling}(2004)}]{Skilling2004AIPC..735..395S}
{Skilling} J (2004) {Nested Sampling}. In: {Fischer} R, {Preuss} R, {Toussaint}
  UV (eds) Bayesian Inference and Maximum Entropy Methods in Science and
  Engineering: 24th International Workshop on Bayesian Inference and Maximum
  Entropy Methods in Science and Engineering, American Institute of Physics
  Conference Series, vol 735, pp 395--405, \doi{10.1063/1.1835238}

\bibitem[{Skilling(2006)}]{skilling2006}
Skilling J (2006) Nested sampling for general bayesian computation. Bayesian
  Anal 1(4):833--859, \doi{10.1214/06-BA127},
  \urlprefix\url{https://doi.org/10.1214/06-BA127}

\bibitem[{{Speagle}(2020)}]{Speagle2020MNRAS.493.3132S}
{Speagle} JS (2020) {DYNESTY: a dynamic nested sampling package for estimating
  Bayesian posteriors and evidences}. \mnras 493(3):3132--3158,
  \doi{10.1093/mnras/staa278}, \eprint{1904.02180}

\bibitem[{{Szab{\'o}} et~al.(2013){Szab{\'o}}, {Szab{\'o}}, {D{\'a}lya},
  {Simon}, {Hodos{\'a}n}, and {Kiss}}]{Szabo2013A&A...553A..17S}
{Szab{\'o}} R, {Szab{\'o}} GM, {D{\'a}lya} G, {Simon} AE, {Hodos{\'a}n} G,
  {Kiss} LL (2013) {Multiple planets or exomoons in Kepler hot Jupiter systems
  with transit timing variations?} \aap 553:A17,
  \doi{10.1051/0004-6361/201220132}, \eprint{1207.7229}

\bibitem[{{Tinetti} et~al.(2018){Tinetti}, {Drossart}, {Eccleston}, {Hartogh},
  {Heske}, {Leconte}, {Micela}, {Ollivier}, {Pilbratt}, {Puig}, {Turrini},
  {Vandenbussche}, {Wolkenberg}, {Beaulieu}, {Buchave}, {Ferus}, {Griffin},
  {Guedel}, {Justtanont}, {Lagage}, {Machado}, {Malaguti}, {Min},
  {N{\o}rgaard-Nielsen}, {Rataj}, {Ray}, {Ribas}, {Swain}, {Szabo}, {Werner},
  {Barstow}, {Burleigh}, {Cho}, {du Foresto}, {Coustenis}, {Decin}, {Encrenaz},
  {Galand }, {Gillon}, {Helled}, {Morales}, {Mu{\~n}oz}, {Moneti}, {Pagano},
  {Pascale}, {Piccioni}, {Pinfield}, {Sarkar}, {Selsis}, {Tennyson}, {Triaud},
  {Venot}, {Waldmann}, {Waltham}, {Wright}, {Amiaux}, {Augu{\`e}res},
  {Berth{\'e}}, {Bezawada}, {Bishop}, {Bowles}, {Coffey}, {Colom{\'e}},
  {Crook}, {Crouzet}, {Da Peppo}, {Sanz}, {Focardi}, {Frericks}, {Hunt},
  {Kohley}, {Middleton}, {Morgante}, {Ottensamer}, {Pace}, {Pearson},
  {Stamper}, {Symonds}, {Rengel}, {Renotte}, {Ade}, {Affer}, {Alard}, {Allard},
  {Altieri}, {Andr{\'e}}, {Arena}, {Argyriou}, {Aylward}, {Baccani}, {Bakos},
  {Banaszkiewicz}, {Barlow}, {Batista}, {Bellucci}, {Benatti}, {Bernardi},
  {B{\'e}zard}, {Blecka}, {Bolmont}, {Bonfond}, {Bonito}, {Bonomo}, {Brucato},
  {Brun}, {Bryson}, {Bujwan}, {Casewell}, {Charnay}, {Pestellini}, {Chen},
  {Ciaravella}, {Claudi}, {Cl{\'e}dassou}, {Damasso}, {Damiano}, {Danielski},
  {Deroo}, {Di Giorgio}, {Dominik}, {Doublier}, {Doyle}, {Doyon}, {Drummond},
  {Duong}, {Eales}, {Edwards}, {Farina}, {Flaccomio}, {Fletcher}, {Forget},
  {Fossey}, {Fr{\"a}nz}, {Fujii}, {Garc{\'\i}a-Piquer}, {Gear}, {Geoffray},
  {G{\'e}rard}, {Gesa}, {Gomez}, {Graczyk}, {Griffith}, {Grodent}, {Guarcello},
  {Gustin}, {Hamano}, {Hargrave}, {Hello}, {Heng}, {Herrero}, {Hornstrup},
  {Hubert}, {Ida}, {Ikoma}, {Iro}, {Irwin}, {Jarchow}, {Jaubert}, {Jones},
  {Julien}, {Kameda}, {Kerschbaum}, {Kervella}, {Koskinen}, {Krijger}, {Krupp},
  {Lafarga}, {Landini}, {Lellouch}, {Leto}, {Luntzer}, {Rank-L{\"u}ftinger},
  {Maggio}, {Maldonado}, {Maillard}, {Mall}, {Marquette}, {Mathis}, {Maxted},
  {Matsuo}, {Medvedev}, {Miguel}, {Minier}, {Morello}, {Mura}, {Narita},
  {Nascimbeni}, {Nguyen Tong}, {Noce}, {Oliva}, {Palle}, {Palmer}, {Pancrazzi},
  {Papageorgiou}, {Parmentier}, {Perger}, {Petralia}, {Pezzuto},
  {Pierrehumbert}, {Pillitteri}, {Piotto}, {Pisano}, {Prisinzano}, {Radioti},
  {R{\'e}ess}, {Rezac}, {Rocchetto}, {Rosich}, {Sanna}, {Santerne}, {Savini},
  {Scandariato}, {Sicardy}, {Sierra}, {Sindoni}, {Skup}, {Snellen}, {Sobiecki},
  {Soret}, {Sozzetti}, {Stiepen}, {Strugarek}, {Taylor}, {Taylor}, {Terenzi},
  {Tessenyi}, {Tsiaras}, {Tucker}, {Valencia}, {Vasisht}, {Vazan}, {Vilardell},
  {Vinatier}, {Viti}, {Waters}, {Wawer}, {Wawrzaszek}, {Whitworth}, {Yung},
  {Yurchenko}, {Osorio}, {Zellem}, {Zingales}, and
  {Zwart}}]{Tinetti2018ExA....46..135T}
{Tinetti} G, {Drossart} P, {Eccleston} P, {Hartogh} P, {Heske} A, {Leconte} J,
  {Micela} G, {Ollivier} M, {Pilbratt} G, {Puig} L, {Turrini} D,
  {Vandenbussche} B, {Wolkenberg} P, {Beaulieu} JP, {Buchave} LA, {Ferus} M,
  {Griffin} M, {Guedel} M, {Justtanont} K, {Lagage} PO, {Machado} P, {Malaguti}
  G, {Min} M, {N{\o}rgaard-Nielsen} HU, {Rataj} M, {Ray} T, {Ribas} I, {Swain}
  M, {Szabo} R, {Werner} S, {Barstow} J, {Burleigh} M, {Cho} J, {du Foresto}
  VC, {Coustenis} A, {Decin} L, {Encrenaz} T, {Galand } M, {Gillon} M, {Helled}
  R, {Morales} JC, {Mu{\~n}oz} AG, {Moneti} A, {Pagano} I, {Pascale} E,
  {Piccioni} G, {Pinfield} D, {Sarkar} S, {Selsis} F, {Tennyson} J, {Triaud} A,
  {Venot} O, {Waldmann} I, {Waltham} D, {Wright} G, {Amiaux} J, {Augu{\`e}res}
  JL, {Berth{\'e}} M, {Bezawada} N, {Bishop} G, {Bowles} N, {Coffey} D,
  {Colom{\'e}} J, {Crook} M, {Crouzet} PE, {Da Peppo} V, {Sanz} IE, {Focardi}
  M, {Frericks} M, {Hunt} T, {Kohley} R, {Middleton} K, {Morgante} G,
  {Ottensamer} R, {Pace} E, {Pearson} C, {Stamper} R, {Symonds} K, {Rengel} M,
  {Renotte} E, {Ade} P, {Affer} L, {Alard} C, {Allard} N, {Altieri} F,
  {Andr{\'e}} Y, {Arena} C, {Argyriou} I, {Aylward} A, {Baccani} C, {Bakos} G,
  {Banaszkiewicz} M, {Barlow} M, {Batista} V, {Bellucci} G, {Benatti} S,
  {Bernardi} P, {B{\'e}zard} B, {Blecka} M, {Bolmont} E, {Bonfond} B, {Bonito}
  R, {Bonomo} AS, {Brucato} JR, {Brun} AS, {Bryson} I, {Bujwan} W, {Casewell}
  S, {Charnay} B, {Pestellini} CC, {Chen} G, {Ciaravella} A, {Claudi} R,
  {Cl{\'e}dassou} R, {Damasso} M, {Damiano} M, {Danielski} C, {Deroo} P, {Di
  Giorgio} AM, {Dominik} C, {Doublier} V, {Doyle} S, {Doyon} R, {Drummond} B,
  {Duong} B, {Eales} S, {Edwards} B, {Farina} M, {Flaccomio} E, {Fletcher} L,
  {Forget} F, {Fossey} S, {Fr{\"a}nz} M, {Fujii} Y, {Garc{\'\i}a-Piquer} {\'A},
  {Gear} W, {Geoffray} H, {G{\'e}rard} JC, {Gesa} L, {Gomez} H, {Graczyk} R,
  {Griffith} C, {Grodent} D, {Guarcello} MG, {Gustin} J, {Hamano} K, {Hargrave}
  P, {Hello} Y, {Heng} K, {Herrero} E, {Hornstrup} A, {Hubert} B, {Ida} S,
  {Ikoma} M, {Iro} N, {Irwin} P, {Jarchow} C, {Jaubert} J, {Jones} H, {Julien}
  Q, {Kameda} S, {Kerschbaum} F, {Kervella} P, {Koskinen} T, {Krijger} M,
  {Krupp} N, {Lafarga} M, {Landini} F, {Lellouch} E, {Leto} G, {Luntzer} A,
  {Rank-L{\"u}ftinger} T, {Maggio} A, {Maldonado} J, {Maillard} JP, {Mall} U,
  {Marquette} JB, {Mathis} S, {Maxted} P, {Matsuo} T, {Medvedev} A, {Miguel} Y,
  {Minier} V, {Morello} G, {Mura} A, {Narita} N, {Nascimbeni} V, {Nguyen Tong}
  N, {Noce} V, {Oliva} F, {Palle} E, {Palmer} P, {Pancrazzi} M, {Papageorgiou}
  A, {Parmentier} V, {Perger} M, {Petralia} A, {Pezzuto} S, {Pierrehumbert} R,
  {Pillitteri} I, {Piotto} G, {Pisano} G, {Prisinzano} L, {Radioti} A,
  {R{\'e}ess} JM, {Rezac} L, {Rocchetto} M, {Rosich} A, {Sanna} N, {Santerne}
  A, {Savini} G, {Scandariato} G, {Sicardy} B, {Sierra} C, {Sindoni} G, {Skup}
  K, {Snellen} I, {Sobiecki} M, {Soret} L, {Sozzetti} A, {Stiepen} A,
  {Strugarek} A, {Taylor} J, {Taylor} W, {Terenzi} L, {Tessenyi} M, {Tsiaras}
  A, {Tucker} C, {Valencia} D, {Vasisht} G, {Vazan} A, {Vilardell} F,
  {Vinatier} S, {Viti} S, {Waters} R, {Wawer} P, {Wawrzaszek} A, {Whitworth} A,
  {Yung} YL, {Yurchenko} SN, {Osorio} MRZ, {Zellem} R, {Zingales} T, {Zwart} F
  (2018) {A chemical survey of exoplanets with ARIEL}. Experimental Astronomy
  46(1):135--209, \doi{10.1007/s10686-018-9598-x}

\bibitem[{{Vanderburg} et~al.(2017){Vanderburg}, {Becker}, {Buchhave},
  {Mortier}, {Latham}, {Charbonneau}, {Lopez-Morales}, and {HARPS-N
  Collaboration}}]{Vanderburg2017AAS...23040204V}
{Vanderburg} A, {Becker} J, {Buchhave} LA, {Mortier} A, {Latham} DW,
  {Charbonneau} D, {Lopez-Morales} M, {HARPS-N Collaboration} (2017) {Precise
  Masses in the WASP-47 Multi-Transiting Hot Jupiter System}. In: American
  Astronomical Society Meeting Abstracts \#230, American Astronomical Society
  Meeting Abstracts, vol 230, p 402.04

\bibitem[{{von Braun} et~al.(2011){von Braun}, {Boyajian}, {ten Brummelaar},
  {Kane}, {van Belle}, {Ciardi}, {Raymond}, {L{\'o}pez-Morales}, {McAlister},
  {Schaefer}, {Ridgway}, {Sturmann}, {Sturmann}, {White}, {Turner},
  {Farrington}, and {Goldfinger}}]{vonBraun2011ApJ...740...49V}
{von Braun} K, {Boyajian} TS, {ten Brummelaar} TA, {Kane} SR, {van Belle} GT,
  {Ciardi} DR, {Raymond} SN, {L{\'o}pez-Morales} M, {McAlister} HA, {Schaefer}
  G, {Ridgway} ST, {Sturmann} L, {Sturmann} J, {White} R, {Turner} NH,
  {Farrington} C, {Goldfinger} PJ (2011) {55 Cancri: Stellar Astrophysical
  Parameters, a Planet in the Habitable Zone, and Implications for the Radius
  of a Transiting Super-Earth}. \apj 740(1):49,
  \doi{10.1088/0004-637X/740/1/49}, \eprint{1106.1152}

\bibitem[{{Weiss} et~al.(2017){Weiss}, {Deck}, {Sinukoff}, {Petigura}, {Agol},
  {Lee}, {Becker}, {Howard}, {Isaacson}, {Crossfield}, {Fulton}, {Hirsch}, and
  {Benneke}}]{Weiss2017AJ....153..265W}
{Weiss} LM, {Deck} KM, {Sinukoff} E, {Petigura} EA, {Agol} E, {Lee} EJ,
  {Becker} JC, {Howard} AW, {Isaacson} H, {Crossfield} IJM, {Fulton} BJ,
  {Hirsch} L, {Benneke} B (2017) {New Insights on Planet Formation in WASP-47
  from a Simultaneous Analysis of Radial Velocities and Transit Timing
  Variations}. \aj 153(6):265, \doi{10.3847/1538-3881/aa6c29},
  \eprint{1612.04856}

\bibitem[{{Weiss} et~al.(2020){Weiss}, {Fabrycky}, {Agol}, {Mills}, {Howard},
  {Isaacson}, {Petigura}, {Fulton}, {Hirsch}, and
  {Sinukoff}}]{Weiss2020AJ....159..242W}
{Weiss} LM, {Fabrycky} DC, {Agol} E, {Mills} SM, {Howard} AW, {Isaacson} H,
  {Petigura} EA, {Fulton} B, {Hirsch} L, {Sinukoff} E (2020) {The Discovery of
  the Long-Period, Eccentric Planet Kepler-88 d and System Characterization
  with Radial Velocities and Photodynamical Analysis}. \aj 159(5):242,
  \doi{10.3847/1538-3881/ab88ca}, \eprint{1909.02427}

\bibitem[{{Wheatley} et~al.(2018){Wheatley}, {West}, {Goad}, {Jenkins},
  {Pollacco}, {Queloz}, {Rauer}, {Udry}, {Watson}, {Chazelas}, {Eigm{\"u}ller},
  {Lambert}, {Genolet}, {McCormac}, {Walker}, {Armstrong}, {Bayliss}, {Bento},
  {Bouchy}, {Burleigh}, {Cabrera}, {Casewell}, {Chaushev}, {Chote},
  {Csizmadia}, {Erikson}, {Faedi}, {Foxell}, {G{\"a}nsicke}, {Gillen},
  {Grange}, {G{\"u}nther}, {Hodgkin}, {Jackman}, {Jord{\'a}n}, {Louden},
  {Metrailler}, {Moyano}, {Nielsen}, {Osborn}, {Poppenhaeger}, {Raddi},
  {Raynard}, {Smith}, {Soto}, and {Titz-Weider}}]{Wheatley2018MNRAS.475.4476W}
{Wheatley} PJ, {West} RG, {Goad} MR, {Jenkins} JS, {Pollacco} DL, {Queloz} D,
  {Rauer} H, {Udry} S, {Watson} CA, {Chazelas} B, {Eigm{\"u}ller} P, {Lambert}
  G, {Genolet} L, {McCormac} J, {Walker} S, {Armstrong} DJ, {Bayliss} D,
  {Bento} J, {Bouchy} F, {Burleigh} MR, {Cabrera} J, {Casewell} SL, {Chaushev}
  A, {Chote} P, {Csizmadia} S, {Erikson} A, {Faedi} F, {Foxell} E,
  {G{\"a}nsicke} BT, {Gillen} E, {Grange} A, {G{\"u}nther} MN, {Hodgkin} ST,
  {Jackman} J, {Jord{\'a}n} A, {Louden} T, {Metrailler} L, {Moyano} M,
  {Nielsen} LD, {Osborn} HP, {Poppenhaeger} K, {Raddi} R, {Raynard} L, {Smith}
  AMS, {Soto} M, {Titz-Weider} R (2018) {The Next Generation Transit Survey
  (NGTS)}. \mnras 475(4):4476--4493, \doi{10.1093/mnras/stx2836},
  \eprint{1710.11100}

\bibitem[{{Winn}(2010)}]{Winn2010exop.book...55W}
{Winn} JN (2010) {Exoplanet Transits and Occultations}, pp 55--77

\bibitem[{{Zingales} et~al.(2018){Zingales}, {Tinetti}, {Pillitteri},
  {Leconte}, {Micela}, and {Sarkar}}]{Zingales2018ExA....46...67Z}
{Zingales} T, {Tinetti} G, {Pillitteri} I, {Leconte} J, {Micela} G, {Sarkar} S
  (2018) {The ARIEL mission reference sample}. Experimental Astronomy
  46(1):67--100, \doi{10.1007/s10686-018-9572-7}, \eprint{1706.08444}

\end{thebibliography}

\end{document}